\documentclass[11pt]{article}
\usepackage{amssymb}
\usepackage{multicol}
\usepackage{amsmath}
\usepackage{epsfig}
\usepackage{longtable}

\begin{document}
\begin{center}
\begin{Large}
{\LARGE\bf Analytic Bias Reduction for $k$--Sample Functionals}
\\[1ex]
\end{Large}
by\\[1ex]
Christopher S. Withers\\
Applied Mathematics Group\\
Industrial Research Limited\\
Lower Hutt, NEW ZEALAND\\[2ex]
Saralees Nadarajah\\
School of Mathematics\\
University of Manchester\\
Manchester M13 9PL, UK
\end{center}
\vspace{1.5cm}
{\bf Abstract:}~~We give analytic methods for nonparametric bias reduction that remove
the need for computationally intensive methods like the bootstrap and the jackknife.

We call an estimate {\it $p$th order} if its bias has magnitude
$n_0^{-p}$ as $n_0 \rightarrow \infty$, where $n_0$ is the sample size (or the minimum sample size
if the estimate is a function of more than one sample).
Most estimates are only first order and require $O(N)$ calculations, where $N$ is
the total sample size.
The usual bootstrap and jackknife estimates are second order but they are
computationally intensive, requiring $O(N^2)$ calculations for one sample.
By contrast Jaeckel's infinitesimal jackknife is an analytic second order
one sample estimate requiring only $O(N)$ calculations.
When $p$th order bootstrap and jackknife estimates are available, they
require  $O(N^p)$ calculations, and so become even more computationally
intensive if one chooses $p>2$.

For general $p$ we provide analytic $p$th order nonparametric estimates that
require only $O(N)$ calculations.
Our estimates are given in terms of the von Mises
derivatives of the functional being estimated, evaluated at the empirical
distribution.

For products of moments an unbiased estimate exists: our form for this
``polykay'' is much simpler than the usual form in terms of power sums.

\section{Introduction and Summary}
\setcounter{equation}{0}
\addtocounter{section}{0}

Let $T (F)$ be any {\it smooth functional} of one or more unknown
distributions $F$ based on random samples from them.
Bias reduction of estimates of $T (F)$, say $T (\widehat{F})$, has been a subject of considerable interest.
Traditionally bias reduction has been based on well known resampling methods like bootstrapping
and jackknifing in nonparametric settings, see Efron (1982).
However, these methods may not be effective in complex situations when the sampling distribution
of the statistic changes too abruptly with the parameter, or when this distribution is very skewed and has heavy tails.
Also the robustness properties of $F$ may not be preserved for $T (F)$ for all $T (\cdot)$.

Recently, various analytical methods have been developed for bias reduction in parametric settings.
Withers (1987) developed methods for bias reduction based on Taylor series expansions.
Sen (1988) established asymptotic normality of
$\sqrt{n} \{ T (\widehat{F}) - T (F) \}$ as $n \rightarrow \infty$
under suitable regularity conditions.
Cabrera and Fernholz (1999, 2004) defined a {\it target estimator}:
for a given $T$ and a parametric family of distributions it is defined by setting
the expected value of the statistic equal to the observed value.
Cabrera and Fernholz (1999, 2004) established under suitable regularity conditions that
the target estimator has smaller bias and mean squared error than the original estimator.
See also Fernholz (2001).

This paper provides the first analytical methods for nonparametric bias reduction.
We give three analytic methods for obtaining unbiased
estimates ({\it UEs}) of any smooth functional $T(F)$.
These UEs are in general infinite series which in practice need to be truncated.
Let us define a {\it $p$th order estimate} as one with bias $O(n_0^{-p})$
as $n_0  \rightarrow \infty$, where $n_0$ is the minimum sample size.
Our truncated $p$th order estimates require only $O(N)$ computations, where $N$ is the total sample size.
By contrast computer intensive methods, like the $p$th
order bootstrap and jackknife estimates require
$O((n_1 \cdots n_k)^p)$ calculations.
Put another way, for fixed $p$, the computational efficiency of
our analytic $p$th order estimate  relative to the $p$th order
bootstrap or jackknife estimate is $O(n_0^{p-1})$.
So, our truncated estimates remove the need for these
computationally intensive methods of nonparametric bias reduction.
The downside is that the details must be worked out for each nonparametric
functional of interest.
This involves calculating the von Mises or functional derivatives of the functional up to order $2p-2$.
When von Mises (1947) introduced these derivatives,
he did not define them uniquely, nor did he give
a method to obtain higher derivatives.
This was rectified in Withers (1983):
the second derivative is {\it not} the derivative of the first derivative,
but requires a `correction' term.
von Mises did give a method for calculating
the first derivative, also known as {\it the influence function} and this is
well known and widely used.
von Mises' expansion for say $T(\widehat{F})$ about ${T}(F)$ was extended
to functionals of more than one distribution in Withers (1988).
This introduced for the first time the partial von Mises derivatives and
showed how to calculate them.

Suppose we observe $k$ independent random samples of sizes $n=(n_1, \ldots, n_k)$ from
 $k$ unknown distributions $F=(F_1, \ldots, F_k)$ on $R^{s_1}, \ldots, R^{s_k}$, where $R$ is the real line.
Let $\widehat{F} = (\widehat{F}_1, \ldots, \widehat{ F}_k)$ be their $k$ empirical (or sample) distributions.
We shall give three $p$th order estimates of any smooth functional
$T(F)$ in terms of the derivatives of $T(F)$ up to order $2p-2$ evaluated at $\widehat{F}$.

As noted we derive these $p$th order estimates from three forms of UE for $ T(F) $.
These are all infinite series {\it unless} $T (F) $ is a polynomial in $F$,
(for example, a polynomial in the moments of $ F_1, \ldots, F_k )$.
Truncation of these series yields three forms of estimates of $T (F)$
of bias $ O(n_0^{-p}) $, where $ n_0 = \min_{i=1}^k n_i $ and $ p \geq 1 $ is any specified integer.

We call these three forms of estimates the
$S, T$ and $V$ estimates.
For $p = 2 $, all three forms of estimates
have $k+1$ terms.
But for $p>2$ the $S$ estimate is the best choice, requiring
fewer terms than the $T$ estimate or the $V$ estimate: see Section 5.
The $T$ estimate is its power series equivalent.
The $V$ estimate is an intermediate form for arriving at the $S$ estimate.

If $T(F)$ is a product of moments or cumulants, then an unbiased estimate
of it exists, and is given by our $S$ estimate with the appropriate choice of $p$.
Special cases include the UEs of the cumulants of Fisher (1929),
the UEs of the central moments of James (1958), and the polykays of
Wishart (1952) given in terms of the power sums
via tables of the symmetric polynomials: see Stuart and Ord (1987, Section 12.22).
Our $S$ estimate gives these polykays in terms of the sample
central moments and so is much more compact and avoids the need for these tables.

For $p=2$ and $k=1$ the relation of our $S$ estimate to the
{\it infinitesimal jackknife} of Jaeckel (1972) is given in Appendix A.
Jaeckel gave formulas for second order estimates in terms of
the derivatives with respect to the weights of $T(\widehat{F})$, where $\widehat{F}$ is now the {\it weighted} empirical distribution.
His formulas are equivalent to our second order one sample $S$ and $T$ estimates.
Our formulas are given in terms
of the second derivatives of $T(F)$.

For the case of {\it one sample} problems
(functionals of only one distribution function), the $S$ and $T$ estimates given here were obtained
by a much more laborious approach in Withers (1994a) starting with an expansion
for $E T(\widehat{F})$ based on a generalised delta method; results were given up to $p=4$.
(This also contains more examples.)
Here these results are extended to $p=6$ for the general {\it $k$-sample} problem.
The present method uses $U$-statistics and so bypasses the need in
Withers (1994a) to differentiate functionals of derivatives.

At this point we give four simple one sample examples dealt with in Section 6:
\begin{itemize}

\item
For univariate data, a second order estimate of the standard deviation
$\ T(F)=\sigma$ is
$T(\widehat{F})\{ 1+m^{-1}s_1(\widehat{F})\}$, where $m=n$ or $n-1$, $s_1(F)=(\beta_4+3)/8$
and $\beta_4$ is the kurtosis, that is the standardised fourth central moment.

\item
For univariate data, a second order estimate of
$\ T (F) =  \mu / \sigma$ is
$T(\widehat{F})+n^{-1}S_1(\widehat{F})$, where $S_1(F)=-\beta_3/2-T(F)(3\beta_4+1)/8$
and $\beta_3$ is the skewness, that is the standardised third central moment.

\item
For bivariate data a second order estimate of the ratio of marginal means $T (F)=\mu_1/\mu_2$, is
\begin{eqnarray*}
(\overline{ X_1} / \overline{X_2}) \{1 + n^{-1}
[\overline{ X_1X_2}/(\overline{X_1} \ \overline{X_2})
-\overline{ X_2^2} / \overline{X_2}^{2} ] \}.
\end{eqnarray*}
(The usual ratio estimate $T(\widehat{F}) = \overline{ X_1} / \overline{X_2}$ has bias $O(n^{-1})$.)

\item
For multivariate data, we give a second order estimate of $T (F) = ( \alpha^\prime \mu)^q$,
where $ \mu $ is the mean vector and
$\alpha$  is any given vector of the same dimension, is
\begin{eqnarray*}
(\alpha^\prime \widehat{\mu})^q\{1-{q \choose 2}(\alpha^\prime \widehat{\mu})^{-2} \alpha^\prime\widehat{V}\alpha/(n-1)\},
\end{eqnarray*}
where $\widehat{\mu}$ is the sample mean, and $\widehat{V}$ is the sample covariance.

\end{itemize}
Applications to skewness and kurtosis have already been given in Withers (1994b).
A $k$-sample univariate example is given by interpreting
$ \mu_i $ in the last example as the mean of the $i$th distribution
sampled, $\widehat{\mu_i}$ as the mean of the $i$th sample, and replacing
$\alpha^\prime\widehat{V}\alpha/(n-1)$ by
$\sum_{i=1}^k \alpha_i^2 \widehat{v_i}/(n_i-1)$, where $\widehat{v_i}$ is the $i$th sample variance.
All these examples allow for a possible initial transformation of the data $ X \rightarrow h (X) $ say.

The three analytic methods for obtaining UEs are as follows.
The simplest UE for $T(F)$ has the {\it $S$ estimate }
\begin{eqnarray}
\sum_{i_{1}, \ldots, i_k =0}^{\infty}
S_{i_1 \cdots i_k} (\widehat{F}) / \{ (n_1 -1)_{i_1} \cdots (n_k -1)_{i_k}\},
\label{1.1}
\end{eqnarray}
where $(m)_i = m(m-1) \cdots (m-i+1) = m! /(m-i)!$.
Clearly this can be transformed if desired to the {\it $T$ estimate}
\begin{eqnarray}
\sum_{i_{1}, \ldots, i_{k} =0}^\infty T_{i_1 \cdots i_k} (\widehat{F}) n_1^{-i_1} \cdots n_k^{-i_k}.
\label{1.2}
\end{eqnarray}
The coefficients $S_{i_1 \cdots i_k} (F)$ and $T_{i_1 \cdots i_k} (F)$
are functions of the partial von Mises derivatives of $T(F)$ of order up to $ ( 2i_1, \ldots, 2i_k )$.
The third form of UE for $T(F)$ has the {\it $V$ estimate}
\begin{eqnarray}
\sum_{r_1, \ldots, r_{k}=0}^{\infty}
\widehat{V}_{r_1 \cdots r_k} / (r_1! \cdots r_k!),
\label{1.3}
\end{eqnarray}
where $V_{r_1 \cdots r_k}$ is determined by the partial derivatives of $T(F)$ of order $ (r_1, \ldots, r_k )$.
If $T(F)$ is a polynomial in $F$ (such as a polynomial in the moments and cumulants  of $F$),
then the $S$ and $V$ forms of the UE reduce to finite sums.

In Section 2 we derive the $V$ estimate (\ref{1.3}) and its multivariate analogue
using $U$-statistics and tables of the symmetric polynomials.
Section 3 derives from it the $S$ estimate (\ref{1.1}).
Section 4 derives the $T$ estimate (\ref{1.2}).
The number of terms required for these estimates and the bootstrap estimate are compared in Section 5.
Finally, Section 6 illustrates the three estimates using various examples, including the four listed above.
We show in particular that our estimates consistently outperform those due to
Sen (1988)  and Cabrera and Fernholz (1999, 2004).
Computer programs in MAPLE for the implementation of the $V$, $S$ and $T$ estimates
for any $p$ and $k$ are given in Appendix B.

We shall often use bold to denote an integer vector, for example,
$\pmb{n}$ for $(n_1,n_2, \ldots, n_k)$, and $\pmb{1}$ for $(1, \ldots, 1)$.
Similarly, we write $\pmb{r!}=r_1! \cdots r_k!$ and $\pmb{(m)_i}=(m_1)_{i_1} \cdots (m_k)_{i_k}$.
With this notation, (\ref{1.1})--(\ref{1.3}) become
\begin{eqnarray*}
\sum_{\pmb{i} = \pmb{0}}^{\infty}  S_{\pmb{i}} (\widehat{F}) /
( \pmb{n} - \pmb{1})_{\pmb{i}}, \ \
\sum_{\pmb {i}=\pmb {0}}^{
\infty} T_{\pmb {i}} (\widehat{F}) /  \pmb{n}^{- \pmb{i}},\ \
\sum_{\pmb {r} = \pmb {0}}^{\pmb{\infty}} \widehat{V}_{\pmb{r}} / \pmb{r} !.
\end{eqnarray*}
We show that the truncated forms
\begin{eqnarray*}
S_{\pmb{np}} (\widehat{F})= \sum_{\pmb{i}=\pmb {0}}^{\pmb{p} -
\pmb{1}} S_{\pmb{i}} (\widehat{F}) / (\pmb{n}-\pmb{1})_{\pmb{i}},\ \
T_{\pmb{np}} (\widehat{F}) = \sum_{\pmb{i} =\pmb{0}}^{\pmb{p} - \pmb{1}}
T_{\pmb{i}} (\widehat{F}) / \pmb{n}^{- \pmb{i}},\ \
V_{\pmb{np}} (\widehat{F}) = \sum_{\pmb{r}=\pmb{0}}^{2 \pmb{p}
- \pmb{2}}  \widehat{V}_{\pmb{r}} / \pmb{r}!,
\end{eqnarray*}
all have bias $O(n_1^{-p_1} + \cdots + n_k^{-p_k})$ as $\pmb{n} \rightarrow  \pmb{\infty}$.

\section{The $V$ Form of Unbiased Estimate}
\setcounter{equation}{0}
\addtocounter{section}{0}

\subsection{One Sample}

Let us first consider the case of {\it one} distribution $F$ on $R^s$ and
one sample $X_1, \ldots, X_n$.
For $G$ any distribution on $R^s$ the
von Mises-Taylor expansion of $T(F)$ about $T(G)$ is
\begin{eqnarray}
T (F) = \sum_{r=0}^\infty V_r (F,G) /r !,
\label{2.1}
\end{eqnarray}
where $V_0 (F,G) = T(G)$ and
$V_r (F,G) = \int \cdots \int T_G (x_1 \cdots x_r) d F(x_1) \cdots dF(x_r)$
for $r \geq 1$, and $T_G (x_1 \cdots x_r)$ is
the $r$th order (von Mises) derivative of $T(G)$, uniquely defined by (\ref{2.1})
subject to the constraints that $T_G (x_1 \cdots x_r)$ is not altered by permuting
$x_1, \ldots, x_r$, and $\int T_G (x_1 \cdots x_r) d G (x_1) =0$.

The first derivative or {\it influence function} of $T(G)$ is just
$ T_G(x)=\lim_{\epsilon \rightarrow0} (T(G_\epsilon)-T(G))/\epsilon$,
where $G_\epsilon(y)=(1-\epsilon)G(y)+\epsilon I(y \leq x)$ and $I(A)$ is $1$
or $0$ for $A$ true or false.
A simple method of obtaining $T_G(x_1 \cdots x_r)$ from $T_G (x_1 \cdots x_{r-1})$ was given in Withers (1983).
For example, $S(G)=T_G(x)$ has derivative $S_G(y)=T_G(x,y)-T_G(y)$ so this gives the
second derivative of $T(G)$ in terms of its first derivative.
Similarly, $S(G)=T_G(x,y)$ has derivative $S_G(z)=T_G(x,y,z)-T_G(z,y)-T_G(x,z)$ so this gives the third derivative.
The analogous formula
holds for the derivative of the general $r$th order derivative, with
$r$ `correction' terms subtracted.

For fixed $G$ an UE of $V_r (F, G)$ is the $U$-statistic
\begin{eqnarray}
V_r^n (\widehat{F}, G) =
\sum_r T_G (X_{i_1} \cdots X_{i_r}) / (n)_r,
\label{2.2}
\end{eqnarray}
where $\sum_r$ sums over all $(n)_r$ permutations of distinct $i_1, \ldots, i_r$ in $1, \ldots, n$.
So,
\begin{eqnarray*}
n V_1^n (\widehat{F}, G)&=& \sum_{i=1}^n T_G (X_{i}) = n \int T_G (x) d \widehat{F} (x),
\\
(n)_2 V_2^n (\widehat{F}, G) &=& \sum_{i\neq j}^n T_G (X_{i}, X_j) = (\sum_{i,j=1}^n - \sum_{i =j =1}^n) T_G (X_{i}, X_j)
\\
&=& n^2 \int \int T_G (x_1, x_2) d \widehat{F}(x_1) d  \widehat{F}(x_2) - n  \int T_G (x_1^2) d \widehat{F}(x_1),
\end{eqnarray*}
where $ T_G (x_1^2)=T_G (x_1, x_1)$, and so on.

Note that $V_r^n (\widehat{F}, G)$ can be written down using the tables of the symmetric
polynomials in Stuart and Ord (1987, Appendix Table 10, page 554) for $ r \leq 6 $
and David and Kendall (1949) for $r \leq 12$.
The last column of these tables expresses the symmetric polynomials
$[1^r] = \sum_r x_{i_{1}} \cdots x_{i_{r}}$, where $\sum_r$  sums over
distinct $i_1, \ldots, i_r$ in $1, \ldots, n$ say,
in terms of the power sum functions
$(j) = \sum_{i=1}^n x_{i}^j$ for $1 \leq j \leq r$.
For example, $[1^4]=-6 (4) +8 (3 1)  + 3(2^2) -6 (2 1^2)  + (1^4)$,
where for $\pi = (\pi_1, \ldots, \pi_m)$, $(\pi) =(\pi_1) \cdots (\pi_m)$.
In general
\begin{eqnarray}
[1^r] = \sum_\pi^r  (\pi) c (\pi),
\label{2.3}
\end{eqnarray}
where $\sum_\pi^r$ sums over all partitions  $\pi$ of $r$, and the numerical
coefficients $c (\pi)$ are provided in the last column of the tables.
The MAPLE procedure {\sf symmpoly(...)} in Appendix B expresses a given number of symmetric polynomials
in terms of power sum functions.

However, the relation (\ref{2.3}) remains true if $[1^r]$ is redefined as
$\sum_r t(x_{i_1}, \ldots, x_{i_r})$ and $(\pi) = (\pi_1 \cdots \pi_m)$ is
redefined as $\sum_{i_{1}=1}^n \cdots \sum_{i_{m} =1}^n  t(x_{i_1}^{\pi_1}, \ldots, x_{i_m}^{\pi_m})$, where
$x_1^{\pi_1}$ denotes $\pi_1$ arguments equal to $x_1$, not a power, and $t$
is an arbitrary symmetric polynomial of $r$ variables.
In particular taking
$x_i \equiv X_{i}$ and $t \equiv T_G$,
$[1^r] =(n)_r V_r^n (\widehat{F}, G)$, and $(\pi) = T_{\widehat{F}G} [\pi] n^{m(\pi)}$,
where $m=m(\pi)$ is the number of
elements in $\pi$, and
\begin{eqnarray}
T_{\widehat{F} G} [\pi] = \int \cdots \int T_G (x_1^{\pi_1} \cdots x_m^{\pi_m})d \widehat{F}(x_1) \cdots d \widehat{F}(x_m).
\label{Tfun}
\end{eqnarray}
So, (\ref{2.3}) can be written
\begin{eqnarray*}
V_r^n (\widehat{F}, G)= (n)_r^{-1} \sum_{\pi}^r c(\pi) T_{\widehat{F}G} [\pi] n^{m(\pi)},
\end{eqnarray*}
and an UE of $T(F)$ is $\sum_{r=0}^\infty V_r^n (\widehat{F}, G) /r!$.
Since $G$ is arbitrary we may now take $G= \widehat{F}$ and set
$\widehat{V}_r = V_{r}^n (\widehat{F},\widehat{F})$,
$\widehat{T}[\pi]= T_{\widehat{F}\widehat{F}} [\pi]$, $T[\pi]= T_{{F}{F}} [\pi]$.
For example, for second order estimates we shall need $\widehat{T} [2] = \int T_{\widehat{F}} (x,x) d \widehat{F} (x)$.

Our expression above for LHS (\ref{2.2}) at $G=\widehat{F}$ now yields the following $V$ form of UE for $T(F)$:
\begin{eqnarray}
\sum_{r=0}^\infty \widehat{V}_r/r!,
\label{2.6}
\end{eqnarray}
where
\begin{eqnarray}
\widehat{V}_r =(n)_r^{-1} {\sum}_{\pi}^{\sim r} c(\pi) \widehat{T} [\pi] n^{m(\pi)},
\label{Vest}
\end{eqnarray}
and ${\sum}_{\pi}^{\sim r}$ is $\sum_{\pi}^r$ excluding partitions containing 1 because of
the constraint $\int T_{\widehat{F}} (x_1 \cdots x_r) d\widehat{F}(x_1) = 0$.
The first few $V_r$ are
\begin{eqnarray*}
\widehat{V}_0 &=& T(\widehat{F}), \,\, \widehat{V}_1=0, \,\, \widehat{V}_2 = -(n-1)^{-1} \widehat{T}[2],
\\
 \widehat{V}_3 &=& 2(n-1)^{-1}_2 \widehat{T}[3],
\\
\widehat{V}_4 &=& (n-1)^{-1}_3  \{ -6 \widehat{T}[4] +3 \widehat{T}[2^2]n \},
\\
 \widehat{V}_5 &=& 2(n-1)^{-1}_4 \{ 24 \widehat{T}[5] -20 \widehat{T}[32]n \},
\\
\widehat{V}_6 &=& (n-1)^{-1}_5 \{ -120 \widehat{T}[6] +90 \widehat{T}[4 2]n +40 \widehat{T}[3^2]
n-15\widehat{T}[2^3] n^2\}.
\end{eqnarray*}

Using the  $O_p(.)$ notation of Mann and Wald (1943),
since $\widehat{V}_{2r-1}$ and $ \widehat{V}_{2r}$ are $O_p(n^{-r})$,
the UE (\ref{2.6}) is $V_{n: 2p-2} (\widehat{F}) + O_p (n^{-p})$, where
\begin{eqnarray*}
V_{n:j} ( \widehat{F}) = \sum_{r=0}^{j} \ \widehat{V}_{r}  /r \ !.
\end{eqnarray*}
Note that $V_{n: 2 p-2} ( \widehat{F} ) $ estimates $T (F) $ with bias $ O(n^{-p})$:
$V_{n:2} (\widehat{F})$ = $T (\widehat{F}) - (n-1)^{-1} \widehat{T}[2]/2$ has bias $O (n^{-2})$,
$V_{n:4} (\widehat{F})$ = $V_{n:2} (\widehat{F})$ + $(n-1)_2^{-1} \widehat{T} [3]/3 + (n-1)_3^{-1} ( -6 \widehat{T} [4] + 3 \widehat{T}[2^2]n ) /24$
has bias $ O(n^{-3})$, and so on.
The MAPLE procedures {\sf Vestsum(...)} and {\sf Vest(...)} in Appendix B
calculate (\ref{Vest}) for any $r$ and hence $V_{n: 2p-2}$ for any $p$,
so estimates of bias of any order can be obtained.

If $T{(F)}$ is a
polynomial in $F$ of degree $p$, for example, $\mu_p (F)$, $\kappa_p (F)$ or $\mu (F)^p$ then
$T_{G} (x_1 \cdots x_r) =0$ for $r>p$ so that $V_{np} (\widehat{F})$ is an UE.

\subsection{More than One Sample}

Now consider the case of $k \geq 1$ distributions, with $k$ samples
$ \{ X_{1j}, \ldots,  X_{n_{j}j} \}, 1 \leq j \leq k$.
The von Mises--Taylor expansion of $T(F)$ about $T(G)$ for $G =(G_1, \ldots, G_k)$ distributions on $R^{s_1}, \ldots, R^{s_k}$ is
\begin{eqnarray}
T{(F)} = \sum_{r_1=0}^\infty \cdots \sum_{r_k=0}^\infty V_{r_1 \cdots r_k}(F, G) / (r_1! \cdots r_k!) = \sum_{\pmb {r}= \pmb {0}}^{\pmb{\infty}} V_{\pmb {r}} (F, G ) / \pmb{r} !,
\label{2.8}
\end{eqnarray}
where
\begin{eqnarray*}
V_{\pmb {r}} (F, G) &=& \int \cdots \int T_G \scriptsize{\left(
\begin{array}{cccccc}
 1 \ 1 & \cdots & 1 & \cdots & k \ k & \cdots  k
\\ x_1  x_2 & \cdots & x_{r_1} & \cdots &  z_1  z_2 & \cdots  z_{r_k}
\end{array} \right)} \\
&&
d F_1(x_1) \cdots d F_1 (x_{r_1}) \cdots d F_1 (z_1) \cdots dF_k (z_{r_k})
\end{eqnarray*}
and
$ T_G  \left(
\begin{array}{c}
a_1 \cdots a_r \\ x_1 \cdots x_r \end{array} \right)$
is the partial (von Mises) derivative, defined
for $a_1, \ldots, a_r$ in $1, \ldots, k$ and $x_i$ in $R^{s_{a_i}}$.
These were introduced in Withers (1988).
They are determined uniquely by (\ref{2.8}) and the two constraints
\begin{eqnarray*}
T_G {\scriptsize{\left( \begin{array}{ccc}
 a_1 &\cdots& a_r
\\ x_1 &\cdots& x_r
\end{array} \right)}}
\end{eqnarray*}
is not altered by permuting columns, and
\begin{eqnarray*}
 \int T_G {\scriptsize{\left( \begin{array}{ccc}
a_1 &\cdots& a_r\\
x_1 &\cdots & x_r
\end{array} \right)}} d F_{G_{a_1}} (x_1) =0.
\end{eqnarray*}
In practice they are determined by adapting the rules given above for the one
distribution derivatives, namely:
$T_G {\scriptsize{\left( \begin{array}{c} a\\ x \end{array} \right)}} $
is just $S_{G_a}(x)$ for $S(G_a)=T(G)$,
$U(G_b)=T_G {\scriptsize{\left( \begin{array}{c} a\\ x
\end{array} \right)}} $
has derivative
\begin{eqnarray*}
U_{G_b}(y)=
T_G {\scriptsize{\left( \begin{array}{cc}
 a   & b
\\ x  & y
\end{array} \right)}}
- T_G {\scriptsize{\left( \begin{array}{c} a\\ x \end{array} \right)}}
\end{eqnarray*}
and similarly for the derivative of the general derivative.

The term $V_{\pmb{0}} (F, G)$ is interpreted as $T(G)$.
For more details
and examples see Withers (1988, 1994a).
For a given $G$, an UE of $V_{\pmb  r} (F,G)$ is
\begin{eqnarray*}
V_{\pmb {r}}^{\pmb {n}} ( \widehat{F}, G) = \sum_{r_{1} \cdots r_{k}}
T_G{\scriptsize{ \left(
\begin{array}{ccc}
1 \cdots 1 &\cdots& k \cdots k
\\ X_{i_{1}1} \cdots X_{i_{r_{1}}1} &\cdots&
X_{j_{1} k} \cdots X_{j_{r_{k}} k} \end{array} \right)}} /
(\pmb {n})_{\pmb {r}},
\end{eqnarray*}
where
$\sum_{r_1 \cdots r_k}$ sums over all $(n_1)_{r_1}$ permutations
of distinct $i_1, \ldots, i_{r_{1}}$ in $\{1, \ldots, n_1\}, \ldots$ and all  $(n_k)_{r_k}$
permutations of distinct $j_{1}, \ldots, \ j_{r_k}$ in $\{ 1, \ldots, n_k \}$.
So,
\begin{eqnarray*}
\sum_{\pmb {r}=\pmb {0}}^{\pmb{\infty}}
V_{\pmb  r}^{n} (\widehat{F},G) / \pmb{r} !
\end{eqnarray*}
is an UE of $T(F)$.
Taking $G=\widehat{F}$,
\begin{eqnarray*}
\sum_{\pmb {r}=\pmb {0}}^{\pmb{\infty}} \widehat{V}_{\pmb {r}} /\pmb{r!}
\end{eqnarray*}
is a  UE of $T(F)$, where $\widehat{V}_{\pmb r} = V_{\pmb{r}}^{\pmb{n}} (\widehat{F}, \widehat{F})$.
For $i_1, \ldots, i_k $ in $ \{ 0, 1 \}, \widehat{V}_{2 \pmb {r-i}} $ is $O_p (n_1^{-r_1} \cdots n_k^{-r_k} )$.
So, $V_{\pmb {n}: 2 \pmb {p}-\pmb {2}} (\widehat{F})$ estimates $T(F)$
with bias $O(n_1^{-p_1} + \cdots + n_k^{-p_k}) = O (n_0^{-p_0})$, where
\begin{eqnarray*}
V_{\pmb {n}:\pmb{j}} (\widehat{F}) = \sum_{\pmb {r}=\pmb {0}}^{\pmb {j}}
\widehat{V}_{\pmb {r}} / \pmb {r}!, \  \  n_0 = \min_{1 \leq i \leq k} n_{i},
p_0 = \min_{1\leq i \leq k} p_i.
\end{eqnarray*}
However for $p_0 >1$, an estimate of $T(F)$ of bias $O(n_0^{-p_0})$ with fewer
terms than $V_{\pmb {n}: \pmb {2} p_0-\pmb{2} }$ is
\begin{eqnarray}
V_{\pmb {n} p_{0}} (\widehat{F}) = \sum_{r_{0}=0}^{2p_0-2} \widehat{V}_{r_{0}} / r_{0} !,
\label{2.12}
\end{eqnarray}
where
\begin{eqnarray*}
\widehat{V}_{r_{0}} = r_{0}!  \sum_{r_1+ \cdots +r_k =r_0} \widehat{V}_{\pmb {r}} / \pmb {r} !.
\end{eqnarray*}
If $T(F)$ is a polynomial of degree $p_1$ in $F_1, \ldots, p_k$ in $F_k$,
then the UE reduces to the finite sum $V_{\pmb {n}:\pmb {p}} (\widehat{F})$, containing
$(p_1 +1) \cdots (p_k +1)$ terms.  This is because $\widehat{V}_{\pmb r} =0$ if
$r_1 > p_1$ or $\cdots $ or $r_k > p_k$.
Set
\begin{eqnarray*}
T_F (i^r j^s \cdots) = \int \int \cdots T_{F} {\scriptsize{\left(
\begin{array}{ccc} i^r & j^s &\\ && \cdots \\ x^r & y^s & \end{array}
\right)}} d F_j (x) d F_j (y) \cdots,
\end{eqnarray*}
where the first column in the integrand stands for $r$ repeated columns of
${i \choose x}$, and similarly for the other columns.
Set
\begin{eqnarray*}
T_{FG} [\pi_1, \ldots, \pi_k]= T_{FG}(1^{\pi_{11}} \cdots 1^{\pi_{m_11}} \cdots 1^{\pi_{1k}} \cdots 1^{\pi_{m_kk}})
\end{eqnarray*}
for $\pi_1 = (\pi_{11} \cdots \pi_{m_{1}1}  ), \ldots, \pi_k = (\pi_{1k} \cdots \pi_{m_k k})$.
(So, $m_k$ is the length of the vector $\pi_k$.)
Set
\begin{eqnarray*}
\widehat{T}[\pi_1 \cdots \pi_k] = T_{\widehat{F}\widehat{F}}[\pi_1 \cdots \pi_k],\
\widehat{T}(i^rj^s \cdots) = T_{\widehat{F}\widehat{F}}(i^rj^s \cdots),\
{T}(i^rj^s \cdots) = T_{FF}(i^rj^s \cdots).
\end{eqnarray*}
Analogous to (\ref{2.6}) we have
\begin{eqnarray}
\widehat{V}_{\pmb  {r}} = (\pmb {n})_{\pmb {r}}^{-1} \sum_{\pi_{1}}^{\sim r_{1}}
\cdots \sum_{\pi_{k}}^{\sim r_{k}} c ( \pi_{1} ) \cdots c ( \pi_{k} )
\widehat{T} [ \pi_{1}, \ldots, \pi_{k} ] n_{1}^{m( \pi_{1})} \cdots n_{k}^{m ( \pi_{k})}.
\label{Vest2}
\end{eqnarray}
The MAPLE procedures {\sf Vestsum(...)} and {\sf Vest(...)} in Appendix B
calculate (\ref{Vest2}) for any $\pmb{r}$ and hence (\ref{2.12}) for any $p_0$,
so estimates of bias of any order can be obtained.
Let $e_i$ be the $i$th  unit vector in $R^k$.
Then the first six $\widehat{V}_{r_0}$ of (\ref{2.12}) are
\begin{eqnarray*}
\widehat{V}_0 &=&T(\widehat{F}),
\nonumber \\
\widehat{V}_1 &=& \sum_{i=1}^k \widehat{V}_{e_i} =0,
\\
\widehat{V}_2 &=& \sum_{i=1}^k \widehat{V}_{2e_i} = - \sum_{i=1}^k T_{\widehat{F}} (i^2) / (n_i - 1),
\end{eqnarray*}
since $\widehat{V}_{e_i+e_j} = 0 $ for $ i \neq j$,
\begin{eqnarray*}
\widehat{V}_3 = \sum_{i=1}^k \widehat{V}_{3e_i} = 2 \sum_{i=1}^k \widehat{T} (i^3)/(n_i - 1)_2,
\end{eqnarray*}
\begin{eqnarray*}
\widehat{V}_4
&=&
\sum_{i=1}^k \widehat{V}_{4e_i} +
{\scriptsize{\left( \begin{array}{c} 4\\2 \end{array} \right)}} \sum_{1 \leq i < j \leq k} \widehat{V}_{2 e_i +2e_j}
\\
&=& \sum_{i=1}^k \{ -6 \widehat{T}(i^4) + 3 \widehat{T}(i^2 i^2) n_i \} / (n_i - 1)_3
+ 6 \sum_{1 \leq i < j \leq k} \widehat{T}(i^2 j^2)/\{(n_i - 1)(n_j -1) \},
\nonumber
\end{eqnarray*}
\begin{eqnarray*}
\widehat{V}_5 &=& \sum_{ {i}=1}^k \widehat{V}_{5e_i} + {\scriptsize{\left(
\begin{array}{c} 5\\2 \end{array} \right)}} \sum_{i \neq  j } \widehat{V}_{3 e_i+2e_j}
\\
&=& \sum_{ {i}=1}^k \{ 24 \widehat{T}(i^5) - 20 \widehat{T}(i^3 i^2)
n_i \} / (n_i - 1)_4
-20 \sum_{i \neq  j } \widehat{T}(i^3 j^2)/ \{(n_i - 1)_2(n_j - 1)\}
\end{eqnarray*}
and
\begin{eqnarray*}
\widehat{V}_6 &=& \sum_{ {i}=1}^{k} \widehat{V}_{6e_{\pmb {i}}} +
{\scriptsize{\left( \begin{array}{c} 6\\2 \end{array} \right)}}\sum_{i \neq
j } \widehat{V}_{4 e_i +2e_j +}{\scriptsize{ \left( \begin{array}{c} 6\\3
\end{array} \right)}}
\sum_{1 \leq i < j \leq k} \widehat{V}_{3 e_i +3e_j}
+ {\scriptsize{\left( \begin{array}{c} 6\\222 \end{array} \right)}}
\sum_{1 \leq i < j <l
\leq k}\widehat{V}_{2 e_{i} +2e_j +2e_l}
\nonumber
\\
&=& \sum_{i=1}^{k} \left\{ -120 \widehat{T}(i^6) + 90 \widehat{T}
(i^4 i^2)n_{i} + 40 \widehat{T}(i^3 i^3) n_{i}
-15 \widehat{T}(i^2 i^2 i^2)n_{i}^2 \right\}/(n_i -1)_5
\nonumber
\\
&&-15 \sum_{i \neq j } \{ -6 \widehat{T}(i^4 j^2)
+ 3 \widehat{T}(i^2i^2 j^2)n_i /  \}  \{(n_i -1)_3(n_j -1)\}
\nonumber
\\
&& + 80 \sum_{1 \leq i < j \leq k } \widehat{T}(i^3 j^3) / \{(n_i-1)_2
(n_j -1)_2 \}
\nonumber
\\
&&- 90 \sum_{1 \leq i < j < l \leq k } \widehat{T}(i^2 j^2l^2) / \{(n_i-1)
(n_j -1)(n_l -1) \}.
\end{eqnarray*}
By (\ref{2.12}) the above formulas give $V_{n4} (\widehat{F}) = \sum_{r=0}^6 \widehat{V}_r/r!$
as an estimate of $T(F)$ with bias $O(n_0^{-4})$, where $n_0 =\min_{1  \leq i\leq k} n_i$.

\section{The $S$ Estimate}
\setcounter{equation}{0}
\addtocounter{section}{0}

Here we derive the $S$ form of UE, (\ref{1.1}), from the $V$ estimate, (\ref{1.3}).

Suppose first that $k=1$, (that is, univariate data).
Then for $r \geq 2, (n)_r \widehat{V}_r$ is a polynomial
in $n=n_1$ of degree $[r/2]$, the integral part of $r/2$, say:
\begin{eqnarray*}
\widehat{V}_r = (n)^{-1}_r \sum_{1 \leq i \leq r/2} U_{r,i} (\widehat{F}) n^{i} =
(n-1)^{-1}_{r-1} \sum_{0 \leq j \leq r / 2-1} U_{r, j+1} (\widehat{F}) n^j
\end{eqnarray*}
for $r \geq 2$.
Writing
\begin{eqnarray}
n^j = \sum_{k=0}^j [n-r+1]_k \ c_{kjr},
\label{3.2}
\end{eqnarray}
where $[a]_k = a (a + 1) \cdots (a + k - 1)$, we obtain
\begin{eqnarray*}
\widehat{V}_r = \sum_{r /2 \leq i < r} V_{ri} (\widehat{F}) / (n-1)_i,
\end{eqnarray*}
where
\begin{eqnarray*}
V_{r i} (F) = \sum_{0 \leq j\leq r / 2 -1 } c_{r-i-1, j r} U_{r, j+1} (F).
\end{eqnarray*}
The MAPLE procedure {\sf cc(...)} in Appendix B
calculates the coefficients $c_{kjr}$ in (\ref{3.2}) for any given $k$, $j$ and $r$.
It follows that the UE is
\begin{eqnarray*}
\widehat{V}_0 + \sum_{r=2}^\infty \widehat{V}_{r} / r! = \sum_{i=0}^\infty S_i (\widehat{F}) / (n-1)_i,
\end{eqnarray*}
where $S_0 (\widehat{F}) = \widehat{V}_0 = T(\widehat{F})$ and
\begin{eqnarray}
S_i (F) = \sum_{r=i+1}^{2i} V_{r i} (F) /r!
\label{3.5}
\end{eqnarray}
for $i \geq 1$.
So,
\begin{eqnarray}
S_{np} (\widehat{F}) = \sum_{i=0}^{p-1} S_i (\widehat{F}) / (n-1)_i
\label{Sest}
\end{eqnarray}
has bias $ O (n^{-p})$.
The MAPLE procedure {\sf Sest(...)} in Appendix B
can be used to calculate (\ref{Sest}) to obtain a bias of any order.
Using (\ref{3.5}), the first few $S_i (F)$ can be shown to be
\begin{eqnarray}
S_1(F)&=& -T[2]/2,
\label{3.6}\\
S_2(F)&=& T[3]/3 +T[2^2] /8,
\label{3.7}\\
S_3(F)&=& -T[4]/4 +3T[2^2]/8 -T[32]/6 -T[2^3]/48,
\label{3.8}\\
S_4(F)&=& T[5]/5 -2T[32]/3 -3T[2^3]/16 +T[42]/8 + T [3^2] /18
\nonumber \\
&& + T[32^2]/24 +T[2^4]/384,
\label{3.9}\\
S_5(F)&=& -T[6]/6 +5T[42]/8 +5T[3^2]/18 -T[52]/10 - T[43] /12 +3 T[2^4]/64
\nonumber \\
&&-T[42^2]/32 - T[3^2 2]/36 - T[32^3]/144 -T[2^5]/3840, \label{3.10}\\
S_6(F)&=& T[7]/7 -3T[52]/5 -T[43]/2+T[32^2]/140 + 127 T [2^4] /64 \nonumber \\
&&-13 T[42^2]/32 -377 T[3^2 2]/1008
+T[62]/12 +T[53]/15 +T[4^2]/32 \nonumber\\
&&- T[32^3]/48 +T[52^5]/40 +T[432]/24
+T[3^3]/324 -T[2^5]/160 \nonumber \\
&&+T[3^2 2^2]/144
+ T[42^3] /192 +T[32^4]/1152 +T[2^6]/46080. \label{3.11}
\end{eqnarray}
So, $ S_{n2} (\widehat{F}) = T(\widehat{F}) - (n-1)^{-1} T[2]/2 $ has bias $ O(n^{-2})$,
$ S_{n3} (\widehat{F}) = S_{n2} (\widehat{F}) + (n-1)_2^{-1} \{ T[3]/3 + T[2^2]/8 \}$ has bias $O (n^{-3})$, and so on.

Now suppose $k\geq 1$, (that is, multivariate data).
The UE
\begin{eqnarray*}
\sum_{\pmb {r}=\pmb {0}}^{\infty} \widehat{V}_{\pmb  r }/
\pmb {r}! = \sum_{r_{0} =0}^{\infty} \widehat{V}_{r_{0}} / r_0!
\end{eqnarray*}
can be written
\begin{eqnarray*}
\sum_{\pmb{i}= \pmb{0}}^{\infty}
S_{\pmb  i } (\widehat{F}) /(\pmb {n} -\pmb {1})_{\pmb  i} = \sum_{i_0=0}^\infty  S_{i_0}^{\pmb  n} (\widehat{F}),
\end{eqnarray*}
where
\begin{eqnarray*}
S_{i_{0}}^{\pmb {n}} (F) = \sum_{\pmb {i}^{'} \pmb {1} = i_{0}}
S_{\pmb {i}}(F)/ (\pmb {n} - \pmb {1})_{\pmb {i}} = O(n_0^{-i_0})
\end{eqnarray*}
for $n_0 = \min_{1\leq i \leq k} n_i$.
An estimate of $T(F)$ of bias $O (n_0^{-p_0})$ is
\begin{eqnarray}
S_{\pmb {n}p_0} (\widehat{F}) = \sum_{i_0=0}^{p_0-1} S_{i_0}^{\pmb  n}(\widehat{F}).
\label{3.13}
\end{eqnarray}
The MAPLE procedure {\sf Sest(...)} in Appendix B
can be used to calculate (\ref{3.13}) to obtain a bias of any order.
For example, to obtain an estimate of bias $O (n_0^{-4})$ one can show that
\begin{eqnarray}
S_0^{\pmb  n} (F) &=& T(F),
\label{3.14}\\
S_1^{\pmb  n} (F) &=& - (1/2) \sum_{i=1}^k T(i^2) / (n_i -1),
\label{3.15}\\
S_2^{\pmb  n} (F) &=& S_{\pmb  n}^2 + S_{\pmb  n}^{11},
\label{3.16}
\end{eqnarray}
where
\begin{eqnarray}
S_{\pmb  n}^2 &=& \sum_{i=1}^k \{ T(i^3)/3 + T(i^2i^2)/8 \} (n_i -1)_2,
\nonumber\\
S_{\pmb  n}^{11} &=& (1/4)\sum_{1\leq i <j \leq k}  T(i^2j^2 )\{ (n_i -1) (n_j-1)\}
\nonumber
\end{eqnarray}
and
\begin{eqnarray}
S_3^{\pmb  n} (F) &=& S_{\pmb  n}^3 + S_{\pmb  n}^{21}+ S_{\pmb  n}^{111},
\label{3.17}
\end{eqnarray}
where
\begin{eqnarray}
S_{\pmb  n}^3 &=& \sum_{i=1}^k \{ -T(i^4)/4 + 3 T(i^2 i^2)/8
-T(i^3 i^2)/6 -T (i^2 i^2 i^2)/48 \} /(n_i -1)_3,
\nonumber\\
S_{\pmb  n}^{21} &=& (-1/2) \sum_{i \neq j} \{ T(i^3 j^2)/3 +T
(i^2 i^2 j^2) /8\} / \{ (n_i -1)_2 (n_j -1) \},
\nonumber\\
S_{\pmb  n}^{111} &=& (-1/2)^3 \sum_{1\leq i< j<l \leq k}  T(i^2 j^2 l^2)
/ \{ (n_i-1) (n_j-1)(n_l-1) \}.
\nonumber
\end{eqnarray}
In general
\begin{eqnarray}
S_r^{\pmb  n} (F) = \sum_{\pi}^r S_{\pmb  n}^\pi,
\label{3.18}
\end{eqnarray}
where for $\pi =(i_1, i_2, \ldots)$, $S_{\pmb  n}^{i_1 i_2 \cdots}$ can be
written down from the formulas for $S_{i_1} (F), S_{i_2} (F), \ldots$: let us write
\begin{eqnarray}
S_i (F) = \sum_{\pi}^{*i} d_{i \pi} T[\pi],
\label{3.19}
\end{eqnarray}
where $\sum_{\pi}^{*i}$ sums over partitions $\pi$ of $\{ i+1, \ldots, 2i\}$.
(Therefore many $d_{i\pi}$ are zero.)
Then
\begin{eqnarray}
S_{\pmb {n}}^{i_1 \cdots i_m} (F) = \sum_{\pmb{\pi}}^{\ast i_1}
\cdots \sum_{\pmb{\pi}}^{\ast i_m} d_{i_{1} \pi_{1}} \cdots d_{i_m \pi_m}
\sum_{j_1 \cdots j_m}^k T(j_1^{\pi_1} \cdots j_m^{\pi_{m}}) / \{ (n_{j_{1}}-1)_{i_{1}} \cdots (n_{j_m}-1)_{i_m} \}, \nonumber
\end{eqnarray}
where $j^\pi = j^{\pi_1} j^{\pi_2} \cdots$ and $\sum_{j_1 \cdots j_m}^k $ sums
over $j_1, \ldots, j_m$ distinct in $1, \ldots, k$ with $j_i < j_{i+1}$ if $\pi_i = \pi_{i+1}$.
For example,
\begin{eqnarray}
S_4^{\pmb  n} (F) = S_{\pmb  n}^4 + S_{\pmb  n}^{31} + S_{\pmb  n}^{22} + S_{\pmb  n}^{211} +S_{\pmb  n}^{1111},
\label{3.20}
\end{eqnarray}
where by (\ref{3.9}),
\begin{eqnarray*}
S_{\pmb  n}^4 &=& \sum_{{i}=1}^k \left\{T(i^5)/5 -2 T (i^3 i^2)/3 -3T(i^2 i^2 i^2)/16\right.
\\
&& \left. + T(i^4 i^4 )/8 + T (i^3 i^3)/18 +T(i^3 i^2 i^2)/24
+ T(i^2 i^2 i^2 i^2)/384 \right \} /(n_i -1)_4,
\end{eqnarray*}
and by (\ref{3.6}), (\ref{3.8}),
\begin{eqnarray*}
S_{\pmb  n}^{31} &=& (-1/2)\sum_{i\neq j} \{-T (i^3 j^2)/4 -3T(i^2 i^2 j^2)/18-T(i^3 i^2 j^2)/6
\\
&& - T(i^2 i^2 i^2 j^2 ) /48 \} / \{ (n_i -1)_3 (n_j -1) \},
\end{eqnarray*}
and by (\ref{3.7}),
\begin{eqnarray*}
S_{\pmb  n}^{22} &&= \sum_{i\leq i< j\leq k} \{T (i^3 j^3)/9 +T(i^2 i^2 j^2j^2)/64 \}/\{(n_i -1)_2 (n_j -1)_2 \}
\\
&&+ 2 (1/3) (1/8) \sum_{i\neq j} T(i^3 j^2 j^2)/\{ (n_i -1)_2 (n_j-1) \},
\end{eqnarray*}
and by (\ref{3.6}), (\ref{3.7}),
\begin{eqnarray*}
S_{\pmb  n}^{211} = (-1/2)^2 \sum_{i \neq j, i\neq l, j < l} \{-T (i^3
j^2 l^2) /3 + T(i^2 i^2 j^2 l^2)/ 8 \}/\{ (n_i -1)_2 (n_j -1) (n_l -1) \}
\end{eqnarray*}
and by (\ref{3.6}),
\begin{eqnarray*}
S_{\pmb  n}^{1111} = (-1/2)^4 \sum_{1 \leq i < j < l < m \leq k} T (i^2
j^2 l^2 m^2)/ \{ (n_i -1) (n_j -1) (n_l -1)(n_m-1) \}.
\end{eqnarray*}
So, (\ref{3.13}), (\ref{3.6})--(\ref{3.11}) provide the $S$-estimate of bias $O(n_0^{-6})$.

\section{The $T$ Estimate}
\setcounter{equation}{0}
\addtocounter{section}{0}

The $T$ form of the UE,
$\sum_{\pmb i = 0}^{\pmb{\infty}} T_{\pmb  i} (\widehat{F})
\pmb{n}^{- {\pmb  i}}$,
is easily derived from the $S$ estimate, but is less useful: its
truncated form cannot be an UE for $T(F)$ a polynomial in $F$, and the number
of components in $T_{\pmb  i} (F)$ rapidly increases over the number in $S_{\pmb  i} (F)$ as $\pmb {i}$ increases.
Since
\begin{eqnarray*}
\prod_{j=1}^i (1-j \epsilon)^{-1} = \sum_{\beta=0}^\infty \epsilon^\beta G_{\beta i},
\end{eqnarray*}
we have
\begin{eqnarray}
G_{\beta i} = \sum_{\alpha_1 + \cdots + \alpha_i = \beta} 1^{\alpha_1} \cdots i^{\alpha_i},
\label{Gbeta}
\end{eqnarray}
and
\begin{eqnarray*}
(n-1)^{-1}_i = \sum_{\alpha=i}^\infty n^{-\alpha} D_{\alpha i},
\end{eqnarray*}
where
\begin{eqnarray}
D_{\alpha i} = G_{\alpha -i,i}.
\label{Dalpha}
\end{eqnarray}
The MAPLE procedures {\sf gbetai(...)} and {\sf dalphai(...)} in Appendix B
calculate (\ref{Gbeta}) and (\ref{Dalpha}), respectively.
Set $D_{\pmb{\alpha}\pmb{i}} = D_{\alpha_1 i_1} \cdots D_{\alpha_k i_k}$, so
$D_{\pmb{i}+ \pmb{\beta},\pmb{i}} = G_{\beta_{1} i_{1}} \cdots G_{\beta_{k} i_{k}}$.
Then
\begin{eqnarray*}
( \pmb{n} - \pmb{1})_{\pmb{i}}^{-1} = \sum_{\pmb{\alpha}=
\pmb{i}}^{\pmb{\infty}}
D_{\pmb{\alpha}\pmb{i}} \pmb{n}^{- \pmb{\alpha}}.
\end{eqnarray*}
So, the UE is
\begin{eqnarray*}
\sum_{\pmb {i} = \pmb {0}}^{\pmb{\infty}}
S_{\pmb  i} (\widehat{F}) /
(\pmb {n} - \ \pmb {1})_{\pmb {i}} = \sum_{\pmb{\alpha}=
\pmb{0}}^{\pmb{\infty}} T_{\pmb{\alpha}} (\widehat{F})
\pmb {n}^{-\pmb{\alpha}} = \sum_{r =0}^\infty \, T_r^{\pmb{n}}
(\widehat{F}),
\end{eqnarray*}
where
\begin{eqnarray*}
T_{\pmb{\alpha}} (F)=\sum_{\pmb{i}=\pmb{0}}^{\pmb{\alpha}}
D_{\pmb{\alpha}\pmb{i}} S_{\pmb{i}}(F)
\end{eqnarray*}
and
\begin{eqnarray*}
T_r^{\pmb {n}} (F) = \sum_{\{ \pmb{\alpha^\prime}\pmb{1} = r\}}
T_{\pmb{\alpha}} (F) \pmb{n}^{-\pmb{\alpha}} = O(n_0^{-r}).
\end{eqnarray*}
Note that
\begin{eqnarray}
T_{\pmb{n} p} (\widehat{F}) = \sum_{r=0}^{p-1} T_r^{\pmb {n}} (\widehat{F})
\label{4.9}
\end{eqnarray}
has  bias $O_p (n_0^{-p})$.
The MAPLE procedure {\sf Test(...)} in Appendix B
can be used to calculate (\ref{4.9}) to obtain a bias of any order.
The first few $\{ T_r^{\pmb  n} (F)\}$ are
\begin{eqnarray*}
T_0^{\pmb  n} (F) = T(F),
\end{eqnarray*}
\begin{eqnarray*}
T_1^{\pmb {n}} (F) = - (1/2) \sum_{i=1}^k T(i^2) n_i^{-1},
\end{eqnarray*}
\begin{eqnarray*}
T_2^{\pmb {n}} (F) &=& \sum_{i=1}^k \{ - T(i^2) /2 + T(i^3)/3 + T(i^2 i^2)/ 8 \} /(n_i -1)_2  \nonumber
\\
&& + (1/4) \sum_{i<j} T(i^2 j^2)/ \{ (n_i -1)(n_j-1) \},
\end{eqnarray*}
\begin{eqnarray*}
T_3^{\pmb {n}} (F) &=& \sum_{\pmb {i}=1}^k \{ - T(i^2) /2 + T(i^3) +
3 T(i^2 i^2) /4  -T(i^4)/4
\nonumber
\\
&&- T(i^3 i^2)/6 -T(i^2 i^2 i^2)/48 \} n_i^{- 3}
\nonumber
\\
&& + \sum_{i\neq j} \{T(i^2 j^2)/4- T(i^3 j^2)/6 -T(i^2 i^2 j^2)/16\}  n_i^{-2}n_j^{-1}
\nonumber
\\
&&- \sum_{i<j<l} T(i^2 j^2 l^2) n_i^{-1} n_j^{-1} n_l^{-1}/8
\nonumber
\end{eqnarray*}
and
\begin{eqnarray*}
T_4^{\pmb {n}} (F) &=& \sum_{i=1}^k \{ - T(i^2) /2 + 7T(i^3)/3 + 25T(i^2 i^2) /8 -3 T(i^4)/2
\nonumber \\
&&- 5T(i^3 i^2)/3 -5 T(i^2 i^2 i^2)/16 +T (i^4 i^4)/8
\nonumber \\
&& + T(i^3 i^3)/18 +T(i^2 i^2 i^2)/24+ T(i^2 i^2 i^2 i^2)/384\}n_i^{-4}
\nonumber\\
&&- (1/2) \sum_{i \neq j} \{- T(i^2 j^2 )/2 +T (i^3 j^2)
\nonumber\\
&& -T (i^4 j^2)/4 -T(i^3 i^2 j^2)/6-T (i^2 i^2 i^2 j^2)/48\}n_i^{-3} n_j^{-1}
\nonumber\\
&&+ \sum_{i<j} \{ T(i^2 j^2)/4 + T(i^3 j^3 )/9  +T (i^3 j^2 j^2)/12 +T(i^2 i^2 j^2 j^2 )/64 \} n_i^{-2} n_j^{-2}
\nonumber\\
&& + \sum_{i \neq j} \{ -T (i^3 j^2 )/6 - T(i^2 i^2 j^2) /
6 \} n_i^{-2} n_j^{-2}
\nonumber\\
&& +\sum_{i\neq j, i \neq l, j<l} \{ -T (i^2 j^2 l^2 )/8 + T (i^3 j^2 l^2 )/12 + T (i^2 i^2 j^2 l^2 )/32 \} n_i ^{-2} n_j^{-1} n_l^{-1}
\nonumber\\
&&+ \sum_{i<j<l<m} T(i^2 j^2 l^2 m^2) n_i^{-1} n_j^{-1} n_l^{-1} n_m^{-1}/16.
\end{eqnarray*}

\section{Number of Computations Required}
\setcounter{equation}{0}
\addtocounter{section}{0}

As $n_0 = \min_{1 \leq i \leq k} n_i \rightarrow \infty$, for fixed
$p$ the estimators of bias $O(n_0^{-p} ), V_{\pmb {n} p} (\widehat{F}) $ of
(\ref{2.12}), $S_{\pmb {n} p} (\widehat{F}) $ of (\ref{3.13}), and $T_{\pmb {n} p}(\widehat{F}) $ of
(\ref{4.9}), all require $O(n_0)$ calculations.
This  is in sharp contrast with bootstrap estimates of
bias $O(n_0^{-p})$, such as that of (1.35) of Hall (1992), which requires
$O(n_0^p)$ calculations for the case $k=1$ given there, and so at least this if
generalized to $k>1$.  (Note that Hall's (1.35) should have a factor
$(-1)^{i+1}$ inserted.)
If one allows $p$ to increase and counts the number of terms needed, then as
shown by Table 5.1, $S_{\pmb {n} p}(\widehat{F})$ requires increasingly fewer
terms than do $V_{\pmb {n} p} (\widehat{F})$ and $T_{\pmb {n} p} (\widehat{F})$.

\noindent
[Tables 5.1--5.3 about here.]

The number of terms in an expression of the form $\sum_{1 \leq i_1 < \cdots < i_m \leq k} a_{\pmb {n}, i_1 \cdots i_{m}}$ is $s_m = (k)_m /m!$.
So, for general $k$ one obtains the results in Table 5.2.
For $k=2, 3$ this gives the results in Table 5.3.

\section{Examples}
\setcounter{equation}{0}
\addtocounter{section}{0}

For $k=1 $ and $p \leq 3$, $S_{n p} (F)$ was given in Withers (1994a).
For $k > 1$, $S_{n_{0} p}(\widehat{F})$ and $T_{n_{0} p} (\widehat{F})$ in Withers (1994a)
differ from $S_{\pmb {n} p} (\widehat{F})$ and $T_{\pmb {n} p} (\widehat{F})$ given here.
All have bias $O(n_0^{-p})$, but the forms given here are more natural.

In this section we go over some of the examples of Withers (1994a) to
obtain $S$ estimates, $S_{\pmb {n} p} (\widehat{F})$, of bias $O(n_0^{-p})$ up to $p=7$.
Recall that for $k=1$ this is given by
\begin{eqnarray}
S_{np} (\widehat{F}) = \sum_{i=0}^{p-1} S_i (\widehat{F}) /(n-1)_i
\label{6.1}
\end{eqnarray}
for $S_{i} (F)$ of (\ref{3.6})--(\ref{3.11}), and for $k>1$ by (\ref{3.13})--(\ref{3.20}).
We begin with two problems estimating a general function of the means of the
distributions after initial given transformations of the data.
When this function of the means is a polynomial of total degree $p_0$ we show that
for both problems $S_{np_{0}} (\widehat{F})$ is an UE.
We then give examples estimating functions of central moments.
We use the convention that repeated indices are summed over their range.
For example, in (\ref{5.2}) below, $ i_{1} \cdots j_{1} \cdots $ are implicitly summed over $ 1, \ldots, t$.

\noindent
{\bf Example 6.1}
$k=1, T(F)=g(\mu)$, where $\mu=\int h d F=E h(X) $ for $ X \sim F$,
where $h: R^s \rightarrow R^t $ and $ g: R^t \rightarrow R$ are given functions.

Assume that the partial derivatives of $g(\mu)$ are finite:
$g_{j_1 \cdots j_r} = \partial_{j_1} \cdots \partial_{j_r} g(\mu)$, where $\partial_j = \partial  /  \partial_{\mu_{j}}$.
Then the general derivative of $T(F)$ is $ T_{F}(x_1 \cdots x_r) = g_{j_{1} \cdots j_r} \mu_{j_1 x_1} \cdots \mu_{j_r x_r}$,
where $\mu_{jx} = \mu_{jF} (x) = h_j (x)  - \mu_j$.
So,
\begin{eqnarray}
T[\pi] = g_{i_1 \cdots i_{\pi_1} j_1 \cdots j_{\pi_{2} \cdots}}
\mu [i_1 \cdots i_{\pi_1}] \mu [j_1 \cdots j_{\pi_2}] \cdots,
\label{5.2}
\end{eqnarray}
where $\mu[i_1 i_2 \cdots] = \int (h (x)-\mu)_{i_1} (h(x)-\mu)_{i_2} \cdots dF(x)$, the joint central moment of $h(X)$.
Substituting into (\ref{3.6})--(\ref{3.11}) gives $S_i (F)$ for $i\leq 6$
so (\ref{6.1}) gives an estimate of bias $O(n^{-7})$.
Now for $i \geq 1, S_{i} (F)$ has the form $\sum_{r=i+1}^{2i}
g_{j_1 \cdots j_r} c_{i j_1 \cdots j_r}$.
So, if $g (\mu)$ is a polynomial of
degree $\pmb {p}$ in $\mu$ (that is of degree $p_i$ in $\mu_i$ for $ 1 \leq i \leq t)$,
then summation may be restricted to $j_1, \ldots, j_r$ with
$\sum_{i=1}^r I (j_i= a) \leq p_a $ for $1 \leq a \leq t$, and
$S_i (F) =0$ for $i\geq |\pmb {p}| = p_1 + \cdots + p_t$ so $S_{n, |\pmb {p}|} (\widehat{F})$ is an UE.
So, (\ref{6.1}), (\ref{3.6})--(\ref{3.11}) gives an UE for  all polynomials $g(\mu)$ of total degree seven or less. $\Box$

\noindent
{\bf Example 6.1.1}
$g (\mu) = N^q$, where $N= \alpha' \mu$ for
some $t$-vector $\alpha$.
If $\pi = (\pi_1, \ldots, \pi_m)$ and $r= \pi_1 + \cdots + \pi_m$ then
$T[\pi]$ = $(q)_r N^q \alpha (\pi_1) \cdots \alpha(\pi_m)$,
where $\alpha (i) = N^{-i} \alpha_{j_1} \cdots \alpha_{j_i} \mu [ j_1 \cdots j_i ]$.
For example, $\alpha (2) = N^{-2} \alpha^\prime \widehat{V} \alpha$, where $V=covar \{h(X)\}$.
So, a second order estimate of $(\alpha^\prime \mu)^q$ is
$(\alpha^\prime \widehat{\mu})^q \{1-{q \choose 2}(\alpha^\prime \widehat{\mu})^{-2} \alpha^\prime \widehat{V} \alpha /(n-1)\}$,
where $\widehat{\mu}, \widehat{V}$ are the sample estimates of $\mu= h(X)$, and $ V$.

\noindent
[Figures 6.1 and 6.2 about here.]

Now suppose $q = t = 2$, $\alpha = (\alpha_1, 1 - \alpha_1)$ and $V = I_2$.
The second order estimate reduces to $\{ \alpha_1 \widehat{\mu_1} + (1 - \alpha_1) \widehat{\mu_2} \}^2 - \{ \alpha_1^2 + (1 - \alpha_1)^2 \}/(n-1)$.
If $h (X)$ is bivariate normal with unit means and the given $V$ then
the target estimator (Cabrera and Fernholz, 1999, 2004) of $(\alpha^\prime \mu)^2$ is
$\{ \alpha_1 \widehat{\mu_1} + (1 - \alpha_1) \widehat{\mu_2} \}^2 - \{ \alpha_1^2 + (1 - \alpha_1)^2 \}/n$.
Sen (1988)'s asymptotic normality provides $\widehat{\omega}$,
the maximum likelihood estimate of $\omega$ given
$(\alpha^\prime \widehat{\mu})^2 \sim N (\omega, (3/n) \{ \alpha_1^2 + (1 - \alpha_1)^2 \}^2 + (4 \omega/n) \{ \alpha_1^2 + (1 - \alpha_1)^2 \})$.
Figures 6.1 and 6.2 compare the performance of these three estimates and those obtained by bootstrapping and jackknifing.
Our estimate has the lowest absolute bias and the lowest mean squared error.
The estimates due to Cabrera and Fernholz (1999, 2004) are so close to ours that they are indistinguishable.

Next suppose $t= \alpha =1$ so
$T(F) = \mu^q$, $\mu = Eh(X)$, $\alpha(i) = \mu^{-i} \mu_{i}$, $\mu_i = E (h (X) -\mu)^i$ and
$T[\pi] = (q)_r \mu^{q-r} \mu_{\pi_1 \cdots} \mu_{\pi_m}$.
Also for $i \geq 1$,
\begin{eqnarray}
S_i (F) = \sum_{r=i+1}^{2i} (q)_r \mu^{q-r} S_{ir},
\label{6.4}
\end{eqnarray}
where $S_{12}$ = $-\mu_2/2$, $S_{23}$ = $\mu_3/3$, $S_{24}$ =  $\mu_2^2 /8$,
$S_{34}$ = $-\mu_4 +3\mu_2^2 /8$, $S_{35}$ = $-\mu_3\mu_2 /6, S_{36} = -\mu_2^3 /48$,
$S_{45}$ = $\mu_5/5-2\mu_3 \mu_2 /3$, $S_{46}$ = $-3 \mu_2^3/16 +\mu_4 \mu_2 /8 + \mu_3^2 /18$,
$S_{47}$ = $\mu_3 \mu_2^2 /48$, $S_{48}$ =  $\mu_2^4 /384$,
$S_{56}$ = $-\mu_6/6+5\mu_4 \mu_2 /8 +5\mu_3^2 /18$, $S_{57}$ = $-\mu_5 \mu_2/10 - \mu_4 \mu_3/12$,
$S_{58}$ = $3 \mu_2 4 \mu_2^2 /32 - \mu_3^2 \mu_2/36$,
$S_{59}$ = $-\mu_3 \mu_2^3 /144$, $S_{5,10}$ = $-\mu_2^5 /3840$,
$S_{67}$ = $\mu_7/7 - 3\mu_5 \mu_2/5 -\mu_4 \mu_3/2 +\mu_3 \mu_2^2 /140$,
$S_{68}$ = $127  \mu_2^4/64 -13\mu_4 \mu_2^2/32 -327\mu_3^2 \mu_2/100+ \mu_6 \mu_2/12 + \mu_5 \mu_3 /15 + \mu_4^2 /32$,
$S_{69}$ = $-7  \mu_3\mu_2^3 /48+\mu_5 \mu_2^2/40 +\mu_4\mu_3 \mu_2/24 + \mu_3^3 /324$,
$S_{6, 10}$ = $-\mu_2^5 /160 + \mu_3^2  \mu_2^2 /144+ \mu_4 \mu_2^3/192$,
$S_{6, 11}$ =  $\mu_3 \mu_2^4 /1152$ and $S_{6, 12}$ = $\mu_2^6 /46080$.
This gives $S_i(F)$ for $i\leq 6$ and so $S_{n 7} (\widehat{F}) $ of bias $O(n^{-7})$.
If $q$ is a positive integer, then $S_i (F) = 0 $ for $i \geq q$
so $S_{nq} (\widehat{F}) $ of (\ref{6.1}), (\ref{6.4})  is an UE of $\mu^q$.
For example, an UE of $\mu^4$ is given by (\ref{6.1}) with
$S_1(F)=-6\mu^2\mu_2$, $S_2(F)=8\mu\mu_3+3\mu_2^2$, $S_3(F)=-6\mu_4+9\mu_2^2$.
This result may be checked by solving the
system of seven equations given by Wishart (1952, page 5).
Alternatively one may follow the method of Stuart and Ord (1987, Section 12.22) using their tables of the symmetric polynomials.
For example, one obtains the UE $T_n(\widehat{F})$ for $\mu^4$, where
\begin{eqnarray*}
(n-1)_3T_n(F)&=&(n^3-8n^2+23n-30)m_4 - n(n^2-7n+4)m_3m_1-n(n^2-6n+6)m_2^2
\\
&&+n^2(n-9)m_2m_1^2+n^3m_1^4,
\end{eqnarray*}
where $m_i=EX^i$.
Clearly our method gives the simpler form.
$\Box$

\noindent
{\bf Example 6.1.2}
$t=2$, $g (\mu)=\mu_1 \mu_2^{-1}$.
Then $g_{12^j} = (-1)^j j ! \ \mu_2 ^{-j-1} = a_j$ say,
$g_{2^{j}} = \mu_1 a_j$, \ and $g_\pi =0$ unless $\pi$ is a permutation of $2^j$ or $12^j$ for some $j$.
For $ \pi =(\pi_1, \pi_2, \ldots)$ set $|\pi| = \pi_1 + \pi_2 + \cdots$
and
\begin{eqnarray*}
\alpha[\pi] = \mu [\pi]/(\mu_{\pi_1} \mu_{\pi_2} \cdots) = E (X_{\pi_1}/\mu_{\pi_1} -1) (X_{\pi_2}/\mu_{\pi_2} -1) \cdots .
\end{eqnarray*}
Then
\begin{eqnarray*}
T[\pi] = \mu_1 \mu_2^{-1} (-1)^{|\pi|} \alpha[2^{\pi_1}] \alpha[2^{\pi_2}]
\cdots \{ |\pi | ! - (|\pi| -1)! \sum_i \pi_i \alpha [12^{\pi_i-1}] /
\alpha [ 2^{\pi_i}]\}.
\end{eqnarray*}
So,
\begin{eqnarray*}
S_1 (F) /T(F) &=& \alpha[12] - \alpha[22] = E X_1 X_2/(\mu_1\mu_2)-E X_2^2/\mu_2^2,
\\
S_2 (F)/T(F) &=&-2 \alpha [222] +2 \alpha[122] + 3 \alpha[22]^2 - 3 \alpha
[12] \alpha[22],
\\
S_3 (F) /T(F) &=& 6 \alpha[12^3] - 6 \alpha[2^4] - 9\alpha [12] \alpha[22]
+ 9 \alpha [22^2]
\\
&&+ 20 \alpha [2^3] \alpha[2^2] -12 \alpha[12^2] \alpha[2^2] -8 \alpha[2^3])\alpha[12]
\\
&&+ 15 \alpha[12] \alpha[22]^2 -15 \alpha[22]^3,
\\
S_4 (F) /T(F) &=& 24 \alpha[12^4] - 24 \alpha[2^5] + 80\alpha [2^3] \alpha[2^2]
- 48 \alpha [12^2] \alpha[2^2]
\\
&&- 32 \alpha [2^3] \alpha[12] +45 \alpha[22^2] (\alpha[12] - \alpha[22])
\\
&&+ 90 \alpha[2^4] \alpha[2^2] -60 \alpha[12^3] \alpha[2^2] -30 \alpha[2^4]\alpha[12]
\\
&&+40\alpha[2^3] (\alpha [2^3] -\alpha[12^2]) - 210 \alpha [2^3] \alpha[2^2]^2
+90 \alpha [12^2] \alpha[2^2]^2
\\
&&+120 \alpha[2^3] \alpha[12] \alpha[22] +105 \alpha[22]^3 (\alpha [22] -
\alpha [12]),
\\
S_5 (F) /T(F) &=& 120\alpha[12^5] - 120 \alpha[2^6] + 150\alpha [2^4]
(3 \alpha [22]-  \alpha[12])
\\
&&- 300 \alpha [12^3] \alpha[2^2] +200\alpha[2^3] (\alpha[2^3] - \alpha[12^2])
\\
&&+ 72\alpha[2^5] (7\alpha[2^3] -3 \alpha[12^3])-360 \alpha[12^4] \alpha[2^3]
\\
&&+60\alpha[2^4] (7\alpha [2^3] -3 \alpha[12^2]) - 240\alpha [12^3] \alpha[2^3]
\\
&&+1890 \alpha [2^2]^3( \alpha[22]- \alpha[12])
+630 \alpha [24]\alpha[22] (\alpha[12]-2 \alpha [22])
\\
&&+630 \alpha [12^3] \alpha [2^2]^2
+210 \alpha[2^3]^2 (\alpha[12]- 4 \alpha[22])
\\
&& -630 \alpha[12^2] \alpha [2^3]
\alpha [22]
+ 945 \alpha [2^2]^4 (\alpha [12] - \alpha [22])
\\
&&+ 840 \alpha[2^2]^2 (3 \alpha [2^3] \alpha[2^2] - \alpha[12^2] \alpha[2^2] -
2\alpha [2^3] \alpha[12]),
\end{eqnarray*}
and similarly for $S_6 (F)$.

\noindent
[Figures 6.3 and 6.4 about here.]

It follows that a second order estimate of $\mu_1/\mu_2$ is
$(\overline{ X_1} / \overline{X_2}) \{1 + n^{-1} [\overline{ X_1X_2}/(\overline{X_1} \ \overline{X_2}) - \overline{ X_2^2} / \overline{X_2}^{2} ] \}$,
where $\overline{ X_1}$ and  $\overline{X_2}$ are the sample means for $X_1$ and $X_2$, respectively.
If $X_i$, $i = 1, 2$ are independent exponential random variables
with means $\mu_i$, $i = 1, 2$ then a target estimator (Cabrera and Fernholz, 1999, 2004) of $\mu_1/\mu_2$ is
$((n_2 - 1)/n_2) (\overline{ X_1} / \overline{X_2})$,
where $n_2$ denotes the sample size for $\overline{X_2}$.
Sen (1988)'s conditions for asymptotic normality do not hold here.
Figures 6.3 and 6.4 compare the performance of the two estimates and those obtained by bootstrapping and jackknifing.
Our estimate again provides the lowest absolute bias and the lowest mean squared error,
but the differences with respect to the target estimator do not appear to be significant.
$\Box$

\noindent
{\bf Example 6.2}
$T(F)= g(\mu)$, where $\mu=(\mu_1, \ldots, \mu_k)$,
$\mu_{i}= \int h_{i} dF_{i} = Eh_{i} (X_{i})$ for
$X_{i} \sim F_{i}, \ h_{i}: R^{s_i} \rightarrow R$
is a given function, and $g: R^k \rightarrow R$ is a given function with
finite partial derivatives $g_{j_1 \cdots j_r}$.
Then
\begin{eqnarray*}
T_F {\scriptsize{\left( \begin{array}{ccc}
 i_1 &\cdots& i_r\\
x_1 &\cdots& x_r \end{array} \right)}} =
g_{i_1  \cdots  i_r} \mu_{i_1 x_1} \cdots \mu_{i_r x_r},
\end{eqnarray*}
where $\mu_{ix} = \mu_{i F_{i}}  (x) = h_i (x) - \mu_i$.
So,
\begin{eqnarray*}
T(i^a j^b \cdots) = g_{i^a j^b \cdots} \mu_a [i] \mu_b [j] \cdots,
\end{eqnarray*}
where $\mu_a [i] = \mu_a (h_{i}(X_i)) = \int (h_i (x) - \mu_i )^a d F_i(x)$.
An estimate of bias $O(n_0^{-p})$ is $S_{\pmb {n}p} (\widehat{F}) = \sum_{i=0}^{p-1} S_{i}^{\pmb {n}} (\widehat{F})$, where by (\ref{3.14})--(\ref{3.20}),
\begin{eqnarray*}
S_0^{\pmb {n}} (F) = g(\mu),
\end{eqnarray*}
\begin{eqnarray*}
S_1^{\pmb {n}} (F) = -(1/2)
\sum_{i=1}^k  g_{i^2} \mu_2 [i] (n_i -1)^{-1},
\end{eqnarray*}
\begin{eqnarray*}
S_2^{\pmb {n}} (F) &=& \sum_{i=1}^k \{g_{i^3} \mu_3 [i] /3 +g_{i^ 4} \mu_2
[i]^2 /8 \} (n_i -1)^{-1}_2
\\
&&+ \sum_{1\leq i < j \leq k} g_{i^2  j^2} \mu_{2} [i]  \mu_2 [j]
(n_i -1)^{-1} (n_j-1)^{-1}/4, \nonumber
\end{eqnarray*}
\begin{eqnarray*}
S_3^{\pmb {n}} (F) &=& \sum_{i=1}^k \{g_{i^4} (- \mu_4 [i] /4 + 3
\mu_2 [i]^2/8)-
g_{i^{5}}  \mu_3 [i] \mu_2 [i]/6 - g_{i^6} \mu_2 [i]^3 /48 \} (n_i-1)^{-1}_3
\\
&&-  \sum_{i\neq j } \{g_{i^3 j^3}  \mu_3[i] \mu_3 [j]/6 +g_{i^4 j^2} \mu_2
[i]^2 \mu_2 [j] /16\} (n_i -1)^{-1}_2 (n_j-1)^{-1}
\\
&& - (1/8) \sum_{i < j < l } g_{i^2 j^2 l^2} \mu_2 [i]\mu_2 [j] \mu_2[l]
 (n_i -1)^{-1} (n_j-1)^{-1} (n_l-1)^{-1}
\end{eqnarray*}
and
\begin{eqnarray*}
S_4^{\pmb {n}} (F) &=& \sum_{i=1}^k \{g_{i^5} ( \mu_5 [i] /5-2 \mu_3 [i]\mu_2[i]/3)
\\
&&+  g_{i^6} ( \mu_4[i]^2/8 + \mu_3 [i]^2/18 -3 \mu_2 [i]^3 /16)
\\
&& + g_{i^7} \mu_3 [i]\mu_2 [i]^2 / 24 + g_{i^8} \mu_2[i]^4/384\}
(n_i -1)_4^{-1} \nonumber
\\
&& -(1/2)  \sum_{i\neq j } \{g_{i^4 j^2} (- \mu_4[i]/4 +3\mu_2 [i]^2/18) \mu_2 [j]
\\
&& - g_{i^5 j^2 } \mu_3 [i]\mu_2 [i] \mu_2 [j] /6 +
g_{i^6 j^2} \mu_2[i]^3 \mu_2[j] /48 \} (n_i -1)^{-1}_3 (n_j-1)^{-1}
\\
&&+ \sum_{i < j } \{g_{i^3j^3 } \mu_3 [i]\mu_3 [j] /9 +g_{i^4 j^4} \mu_2
[i]^2 \mu_2 [j]^2 /64\} (n_i -1)^{-1}_2 (n_j-1)^{-1}_2
\\
&& + \sum_{i \neq j} g_{i^3 j^3} \mu_3
[i] \mu_3 [j] (n_i -1)^{-1}_2 (n_j-1)^{-1}_2/12
\\
&&+ \sum_{i \neq j, i\neq l, j<l} \{g_{i^3 j^2 l^2} \mu_3 [i] \mu_2 [j] \mu_2[l]/12
\\
&&+g_{i^4 j^2 l^2} \mu_4 [i]^2 \mu_2 [j] \mu_2 [l]/32 \} (n_i -1)^{-1}_2
(n_j-1)^{-1} (n_l -1)^{-1}
\\
&&+ \sum_{i <j<l<m} g_{i^2 j^2 l^2 m^2} \mu_2
[i] \mu_2 [j]\mu_2 [l] \mu_2[m]  (n_i -1)^{-1} (n_j-1)^{-1}(n_l-1)^{-1} (n_m-1)^{-1} /16.
\nonumber
\end{eqnarray*}
As in Example 6.1, for $i\geq 1, S_i^{\pmb{n}} (F)$ has the form
$\sum_{r=i+1}^{2i} g_{j_1 \cdots j_r} c_{i j_1 \cdots j_r}^{\pmb{n}}$.
So, if $g(\mu)$ is a polynomial of degree $\pmb{p}$ in $\mu$ (that is degree $p_i$ in $\mu_i$ for
$1 \leq i \leq k)$, then $S_{\pmb{n} |\pmb{p}|} (\widehat{F})$ is an UE of $g(\mu)$ for $|\pmb{p}| = \pmb{p}' \pmb{1}$.
$\Box$

\noindent
{\bf Example 6.2.1}
$g(\mu) = N^q$, where $N=\alpha' \mu$ for some $k$-vector $\alpha$.
Then
\begin{eqnarray*}
g_{i_1 \cdots i_r} = (q)_r N^{q-r} \alpha_{i_1} \cdots \alpha_{i_r}.
\end{eqnarray*}
So,
\begin{eqnarray*}
T(i^a j^b \cdots) = (q)_r N^{q-r} \alpha_a [i] \alpha_b [j] \cdots,
\end{eqnarray*}
where $r = a + b + \cdots$, $\alpha_a [i] = \alpha_i^a \mu_a [i]$, and  for $i \geq 1$,
\begin{eqnarray*}
 S_i^{\pmb{n}} (F) = \sum_{r=i+1}^{2i} (q)_r N^{q-r} S_{ir}^{\pmb{n}},
\end{eqnarray*}
where
\begin{eqnarray*}
S_{12}^{\pmb{n}} = -\sum_{i=1}^k \alpha_2 [i] (n_i -1)^{-1}/2,
\end{eqnarray*}
\begin{eqnarray*}
S_{23}^{\pmb{n}} =  \sum_{i=1}^k \alpha_3 [i] (n_i -1)_2^{-1}/3,
\end{eqnarray*}
\begin{eqnarray*}
S_{24}^{\pmb{n}} =  \sum_{i=1}^k \alpha_2 [i]^2 (n_i -1)_2^{-1} /8
+  \sum_{i<j} \alpha_2 [i] \alpha_2 [j] (n_i-1)^{-1} (n_j-1)^{-1}/4,
\end{eqnarray*}
\begin{eqnarray*}
S_{34}^{\pmb{n}} = \sum_{i=1}^k (- \alpha_4 [i]/4 + 3 \alpha_2[i]^2/8)
(n_i -1)_3^{-1},
\end{eqnarray*}
\begin{eqnarray*}
S_{35}^{\pmb{n}} = -\sum_{i=1}^k \alpha_3 [i] \alpha_2 [i]
(n_i -1)_3^{-1}/6,
\end{eqnarray*}
\begin{eqnarray*}
S_{36}^{\pmb{n}} &=& -\sum_{i=1}^k \alpha_2 [i]^3 (n_i -1)_3^{-1} /48
\\
&&- \sum_{i \neq j} (\alpha_3 [i] \alpha_3 [j]/6 + \alpha_2 [i]^2 \alpha_2
[j] /16) (n_i-1)^{-1}_2 (n_j-1)^{-1}\nonumber
\\
&&-  \sum_{i<j<l} \alpha_2 [i] \alpha_2 [j] \alpha_2 [l] (n_i -1)^{-1}
(n_j-1)^{-1} (n_l -1)^{-1}/8,\nonumber
\end{eqnarray*}
\begin{eqnarray*}
 S_{45}^{\pmb{n}} &=& \sum_{i=1}^k ( \alpha_5 [i]/5 - 2 \alpha_3[i]
\alpha_2[i]/3) (n_i -1)_4^{-1},
\end{eqnarray*}
\begin{eqnarray*}
S_{46}^{\pmb{n}} &=& \sum_{i=1}^k ( \alpha_4 [i]^2/18 + \alpha_3[i]^2 /18-3
\alpha_2[i]^3/16) (n_i -1)_4^{-1}
\\
&&-  \sum_{i\neq j} (- \alpha_4 [i] /4 +3\alpha_2 [i]^2 /18) \alpha_2
[j] (n_i -1)_3^{-1} (n_j-1)^{-1}/2
\\
&&+\sum_{i<j} \alpha_3 [i] \alpha_3 [j] (n_i -1)^{-1}_2 (n_j-1)_2^{-1}/9,
\end{eqnarray*}
\begin{eqnarray*}
S_{47}^{\pmb{n}} &=& \sum_{i=1}^k  \alpha_3 [i] \alpha_2 [i]^2
(n_i -1)_4^{-1} / 384
\\
&&+ \sum_{i\neq j} \alpha_3 [i] \{\alpha_2 [i] \alpha_2[j]
 (n_i -1)_3^{-1} (n_j-1)^{-1}+ \alpha_2[j]^2 (n_i-1)^{-1}_2
(n_j-1)_2^{-1} \} / 12
\\
&&+\sum_{i\neq j, i \neq l, j<l} \alpha_3 [i] \alpha_2 [j]
\alpha_1[l] (n_i -1)^{-1}_2 (n_j-1)^{-1}_2 (n_l-1)_2^{-1}/12
\end{eqnarray*}
and
\begin{eqnarray*}
S_{48}^{\pmb{n}} &=& \sum_{i=1}^k \alpha_3 [i]^4  (n_i -1)_4^{-1}/384
\\
&&+ \sum_{i\neq j} \alpha_2 [i]^3 \alpha_2 [j]
 (n_i -1)_3^{-1} (n_j-1)^{-1}/96
 \\
&&+\sum_{i < j} \alpha_2 [i]^2 \alpha_2 [j]^2(n_i-1)^{-1}_2 (n_j-1)_2^{-1}/64
\\
&&+\sum_{i\neq j, i\neq l, j<l} \alpha_4[i] \alpha_2[j] \alpha_2[l]
(n_i -1)^{-1}_2 (n_j-1)^{-1} (n_l-1)^{-1}/32
\\
&&+\sum_{i<j<l <m } \alpha_2[i] \alpha_2[j]\alpha_2 [l] \alpha_2[m]
(n_i -1)^{-1} (n_j-1)^{-1} (n_l-1)^{-1}(n_m-1)^{-1}/16.
\end{eqnarray*}
For example, a second order estimate of $(\alpha^\prime \mu)^q$ is
\begin{eqnarray*}
(\alpha^\prime \widehat{\mu})^q \{1-{q \choose 2}(\alpha^\prime \widehat{\mu})^{-2}
\sum_{i=1}^k \alpha_i^2 \widehat{v}_i /(n_i-1)\},
\end{eqnarray*}
where $\widehat{\mu_i}, \widehat{v}_i$ are the sample estimates of $\mu_i, v_i=\mu_2[i]$, the mean and variance of $h_i(X_i)$.
$\Box$.

\noindent
{\bf Example 6.2.2}
$k=2$, $g(\mu) = \mu_1 \mu_2 /(\mu_1 + \mu_2)$.
Set $D=\mu_1 + \mu_2$.
Then
$g(\mu) = 1 - \mu_2^2 D^{-1}$ so $g_{1^{a}} \mu_2^{-2} = g_{2^{a}} \mu_1^{-2} = (-D)^{-a-1} a!$.
Also since $\mu_2^2 = (D- \mu_1)^2$,
\begin{eqnarray*}
g_{1^{a} 2^{b}} = -(-D)^{-a-b-1} (a+b-2) ! \{(b^2-b) \mu_{1}^2 - 2ba \mu_1 \mu_2 + (a^2-a) \mu_2^2 \}.
\end{eqnarray*}
So,
\begin{eqnarray*}
S_1^{\mathbf {n}}(F)&=& T(F)^2 D^{-1} \sum_{i=1}^{2} \mu_{i}^{-2} \mu_2
[i] (n_i -1)^{-1},
\\
S_2^{\pmb{n}} (F) &=& T(F)^2 \sum_{i=1}^2 (2D^{-2} \mu_3 [i] - 3 D^{-3}
\mu_2[i]^2) \mu_i^{-2} (n_i -1)^{-1}_2
\\
&&+ D^{-5} (\mu_1^2 + \mu_2^2- 4 \mu_1 \mu_2)\mu_2 [1] \mu_2 [2]
(n_1 -1)^{-1} (n_2-1)^{-1},
\nonumber
\\
S_3^{\pmb{n}} (F) &=& T(F)^2 \sum_{i=1}^2 \mu_i^{-2}\{D^{-3} ( 6\mu_4 [i] -9 \mu_2[i]^2)
\\
&&+20 D^{-4} \mu_3[i]  \mu_2[i] +15D^{-5}  \mu_2 [i]^3 \}
(n_i -1)^{-1}_3 \nonumber
\\
&&- 48 D^{-7} (\mu_1^2+\mu_2^2 - 3 \mu_1 \mu_2) \mu_3 [1] \mu_3 [2] \{
(n_1 -1)_2^{-1} (n_2 -1)^{-1} \nonumber
\\
&&+ (n_1-1)^{-1} (n_2 -1)_2^{-1} \}\nonumber
\\
&&- 3D^{-7} (\mu_1^2 - 8 \mu_1 \mu_2 + 6\mu_2^2) \mu_2 [1]^2 \mu_2 [2]
(n_1-1)_2^{-1} (n_2-1)^{-1}\nonumber
\\
&&- 3D^{-7} (6\mu_1^2 - 8 \mu_1 \mu_2 + \mu_2^2) \mu_2 [2]^2 \mu_2 [1]
(n_2-1)_2^{-1} (n_1-1)^{-1},\nonumber
\end{eqnarray*}
and so on.  $\Box$

\noindent
{\bf Example 6.3}
$k = 1$, $T(F) = \mu_r = \mu_r (h(X)) = E(h(X)-\mu)^r$, where
$\mu=Eh(X)$ for $X \sim F$ and $h: R^s \rightarrow R$ a given function.
By (5.6) of Withers (1994a), its general $p$th  order derivative is
\begin{eqnarray*}
\mu_{rF} (x_1 \cdots x_{p}) = (-1)^{p} \{ (r)_p \mu_{r-p} - (r)_{p-1}
\sum_{i=1}^p (h_{i}^{r-p} - \mu_{r-p+1} h_{i}^{-1}) \} \prod_{j=1}^p h_j,
\end{eqnarray*}
where $h_{i} = \mu_F (x_{i}) = h(x_{i}) -\mu$.
So,
$ T[i_1 i_2 \cdots] = T (1^i 1^{i_2} \cdots)$ and $S_i(F)$ for $i\leq 3$ are
as given there.
For example,
\begin{eqnarray*}
T[i] &=& (-1)^i \{ (r)_{i}\mu_{r- i} \mu_{i} -i
(r)_{i-1} (\mu_{r} - \mu_{r-i+1} \mu_{i-1})\},
\\
 T[4] &=& (r)_4 \mu_{r-4} \mu_4 -4 (r)_3 (\mu_r - \mu_{r-3}\mu_3),
\\
 T[5] &=& -(r)_5 \mu_{r-5}\mu_5 + 5(r)_4 (\mu_r - \mu_{r-3}
\mu_3),
\\
 T[42] &=& (r)_6 \mu_{r-6} \mu_4\mu_2 -2(r)_5 \{-2\mu_{r-5}  \mu_3
\mu_2 + \mu_{r-4} \mu_4 + 2 \mu_{r-2} \mu_2\},
\\
 T[3^2] &=& (r)_6 \mu_{r-6} \mu_3^2- 6 (r)_5 ( \mu_{r-3} \mu_{r-5}\mu_2)\mu_3,
\\
T[32^2] &=& -(r)_7 \mu_{r-7} \mu_3 \mu_2^2+ (r)_6 \{-3\mu_{r-6}\mu_2^3 + 4 \mu_{r-5} \mu_3 \mu_2+ 3 \mu_{r-4} \mu_2^2\},
\\
T[2^4] &=& (r)_8 \mu_{r-8} \mu_2^4-8(r)_7 \mu_{r-6} \mu_2^3.
\end{eqnarray*}
So,
\begin{eqnarray*}
S_4 [F] = \sum_{i=0}^8 \mu_{r-i} (r)_i S_{4-i},
\end{eqnarray*}
where $S_{4-0} = (r)_4$, $S_{4-1} = 0$, $S_{4-2} = -r(r-2)_2 \mu_2/2$, $S_{4-3} = -r (r-3) \mu_3 /3$,
$S_{4-4} = -r \mu_4 + r^2  \mu_2^2/8$, $S_{4-5} = -\mu_5/5+ (r+4) \mu_3 \mu_2 /6$,
$S_{4-6} = -3 \mu_2^3/16 +\mu_4 \mu_2/8+ \mu_{3}^{2} /18 -r \mu_2^3/48$, $S_{4-7} = -\mu_3\mu_2^2/24$ and $S_{4-8} = \mu_2^4/384$.
The fifth order $S$ estimate is now given by (\ref{6.1}).
Also by Example 5.3 of Withers (1994a) one can now obtain an UE for $\mu_r $ for $r \leq 9$.
Obtaining $S_5, S_6$ from (\ref{3.10})--(\ref{3.11}) one can obtain an UE for $\mu_r$ for $r\leq 12$.
Our results agree with those of James (1958, page 6):
by a different method he obtained an UE for $\mu_r $ for $r \leq 6$.
$\Box$

For $T(F)$ a function of a central moment, one may apply the chain rule, as
illustrated in the next example for the case $T(F)=\sigma=\mu_2^{1/2}$.

\noindent
{\bf Example 6.4}
Here we specialise an application of the chain rule given in Appendix A of Withers (1994a).
Suppose $k=1$, $g: R \to R $ is a function with
finite derivatives and $T (F) = g (S(F))$, where $S (F)$ is a smooth univariate functional.
Set $g_r = g^{(r)} (S(F))$.
The third order $S$ estimate of $ T(F) $ is given by (\ref{6.1}), (\ref{3.6}), (\ref{3.7}) in terms of
\begin{eqnarray*}
T [2] & =& g_2 S (1, 1) + g_1 \ S [2],
\\
T [3] & =& g_3 S (1, 1, 1) + 3 g_2 S (1, 1^2) + g_1 \ S [3],
\\
T [2^2] &=& g_4 S (1, 1)^2 \ + g_3 \{ 2 S (1, 1) S [2] + 4 S (ab, a, b) \}
\\
&& + g_2 \{ 4 S (a, ab^2) + S [2]^2 + 2 S (ab, ab) \} + g_1 \ S [2^2],
\end{eqnarray*}
where
\begin{eqnarray*}
S (1, 1) = \int \ S_x^2,\ \
S (1, 1, 1) = \int \ S_x^3,\ \
S(1, 1^2) = \int S_{x} \ S_{xx},\ \
S (ab, a, b) = \int \int \ S_{xy} \ S_x  \  S_y,
\end{eqnarray*}
\begin{eqnarray*}
S (a, ab^2) = \int \int  \ S_x \ S_{xyy},\ \
S (ab, ab) = \int \int \ S_{xy}^2,\ \
S_{x y \cdots} = S_F \ (x, y, \ldots)
\end{eqnarray*}
and integration is with respect to $ F(x)$ and $F(y)$.

Take $ T (F) = \sigma $, \ that is $ S(F) = \mu_2 $ and $ g (s)= s^{1/2}$.
Then $g_r = (1/2)_r \mu_2^{1/2-r}$, where $(r)_i = r! / (r-i)!$,
$\mu_x = x- \mu$, $\mu_{2x} = \mu_x^2 - \mu_2$, $\mu_{2xy} = -2 \mu_x \mu_y$, and higher derivatives vanish.
So, $S (1, 1) = \mu_4 - \mu_2^2$, $S [2] = -2 \mu_2$,
$S (1, 1, 1) = \mu_6 - 3 \mu_4 \mu_2 + 2 \mu_2^2$,
$S (1, 1^2) = -2 (\mu_4 - \mu_2^2)$,
$S [3] = S (a, ab^2) = S [2^2] = 0$,
$S (ab, a, b) = -2 \mu_3^2$ and $S (ab, ab) = 4 \mu_2^2$.
Setting $ \beta_r = \mu_2 \ \mu_2^{-r/2}$, we obtain
$S_1 (F) / \mu_2^{1/2} = (\beta_4 + 3) / 8$ and
$S_2 (F) / \mu_2^{1/2} = (16 \beta_6 + 22 \ \beta_4 + 164 - 15 \beta_4^2) /128$.

\noindent
[Figures 6.5 and 6.6 about here.]

It follows that a second order estimate of $\sigma$ is $\widehat{\sigma} \{ 1 + (1/(n - 1)) (1/8) (\widehat{\beta_4} + 3) \}$.
Suppose now $F$ is exponential with mean $\sigma$.
Then a target estimator (Cabrera and Fernholz, 1999, 2004) of $\sigma$ is the sample mean $\overline{X}$.
Sen (1988)'s asymptotic normality provides $\widehat{\omega}$,
the maximum likelihood estimate of $\omega$ given $\widehat{\sigma} \sim N (\omega, \omega^2/n)$.
Figures 6.5 and 6.6 compare the performance of these three estimates and those obtained by bootstrapping and jackknifing.
Our estimate again gives the lowest absolute bias and the lowest mean squared error.
\hfill $\Box$

We end this section with examples of functions of more than one moment.
These appeal to some results given in Withers (1994a).

\noindent
{\bf Example 6.5}
$k=1, T(F) =g(\mu, \mu_2)$ for $\mu = E Y, \mu_2 = E(Y-\mu)^2,
Y=h(X)$, given $h: R^s \rightarrow R $  and $ g: R^2 \rightarrow R$ with finite derivatives
$g_{j_1 \cdots j_r} = \partial_{j_1} \cdots \partial_{j_r} g(U_1, U_2)$
at $U_1 = \mu$, $U_2= \mu_2$, where $\partial_i = \partial / \partial U_i$.
We give the fourth order $S$ estimate for general $g$, then specialise.
Note that $S_1(F)$, $S_2(F)$, $S_3(F)$ are given by (\ref{3.6})--(\ref{3.8}) and (A14)--(A15),
(A20)--(A23)  of Withers (1994a).
Set
\begin{eqnarray*}
U_{ij ..} (a^I b^J ..,
a^K b^L .., ..) = \int \int .. U_{iF} (a^I b^J ..) U_{jF}
(a^K b^L ..) .. dF(a) d F (b) ...
\end{eqnarray*}
Allowing for permutations the nonzero terms needed are as follows:
\begin{itemize}

\item
for $T[2]$: at $(a^2), U_2 = -2\mu_2$;
at $ (a, a), U_{11} = \mu_2, U_{12} = \mu_3, U_{22} = \mu_4-\mu_2^2$;

\item
for $T[3]$: at $(a, a^2),
U_{12} =  -2 \mu_3,  U_{22} = -2 (\mu_4-\mu_2^2)$;
at $(a, a, a)$, $U_{111} = \mu_3$, $U_{112} = \mu_4-\mu_2^2$, $U_{122} = \mu_5-2 \mu_3 \mu_2$,
$U_{222} = \mu_6 - 3 \mu_4 \mu_2 +2 \mu_2^3$;

\item
for $T[2^2]: U_{22} (ab, ab)=4 \mu_2^2$; at ($ab, a, b)$,
$U_{211} = -2 \mu_2^2$, $U_{212} = U_{221} = -2 \mu_2 \mu_3$, $U_{222}= -2 \mu_3^2$;

\item
for $T[4]: U_{22} (a^2, a^2)= 4 \mu_4$; at  $(a, a, a^2)$,
$U_{112} = \mu_4 - \mu_2^2$, $U_{122} = -2 (\mu_5 - \mu_3 \mu_2)$, $U_{222} =-2 (\mu_6 -2 \mu_4 \mu_2 + \mu^2_2)$;
at $(a, a, a, a)$, $U_{111} = \mu_4$, $U_{1112} = \mu_5 - \mu_3 \mu_2$, $U_{1122} = \mu_6-2 \mu_4 \mu_2 + \mu_2^2$,
$ U_{1222} = \mu_7-3 \mu_5 \mu_2 +3 \mu_3 \mu_2^2$,  $U_{2222} = \mu_8 - 4 \mu_6 \mu_2 +6 \mu_4 \mu_2^2 - 3 \mu_2^4$.

\item
By (A14), $ T[32]=T(a^2 b^3) = \sum_{k=1}^5 \nu_{23.k}$ say, where $\nu_{23.1} = \nu_{23.2} =0$.
For $\nu_{23.3}$: at $(a, ab, b^2)$, $U_{122} = 4\mu_3 \mu_2$, $U_{222} = 4 \mu_3^2$;
at $ (b,ab, ab)$, $U_{122} = 4 \mu_3 \mu_2$, $U_{222} = 4 (\mu_4 - \mu_2^2) \mu_2$.
For $\nu_{23.4}$: at $(ab, a, b, b)$, $U_{2111} = -2\mu_3 \mu_2$,
$U_{2112} = -2 \mu_2 (\mu_4 -\mu_2^2)$, $U_{2122}=-2 \mu_2(\mu_5 -2 \mu_3 \mu_2)$,
$U_{2212} = -2 \mu_3 (\mu_4 - \mu_2^2)$, $U_{2222} = -2 \mu_3 (\mu_5 -2 \mu_3 \mu_2)$.

\item
By (A15), $T[2^3] = T(a^2 b^2 c^2) = \sum_{k=1}^6  \nu_{222.k}$, where $\nu_{222.1} = \nu_{222.2} =0$.
For $\nu_{222.3} = g_{ijk}  B_3^{ijk}$: $U_{222} (ab, bc, ca) = -8 \mu_2^3$.
For $\nu_{222.4} = g_{ijkl}  B_3^{ijkl}$: at $(a, b, ab)$,
$U_{112} = -2 \mu_2^2$, $U_{122} = -2 \mu_3 \mu_2$, $U_{222} = -2 \mu_3^2$;
at $(a, b, ac, bc)$, $U_{1122} =4 \mu_2^3$, $U_{1222} = 4 \mu_3 \mu_2^2$, $U_{2222} = 4 \mu_3^2 \mu_2$.

\end{itemize}
Other components are covered by the above $U_{ij \cdots}$.
$\Box$

\noindent
{\bf Example 6.5.1}
$g(\mu, \mu_2) = \mu \mu_2^{-1/2} = \beta$ say.
Let $\beta_r = \mu_2^{-r/2} \mu_r$.
Then $S_1(F), S_2(F), T[4] = T(1^4)$
and $T[2^2] = T(1^2 1^2)$ are given by Example 5.6 of Withers (1994a).
So, $S_3 (F)$ is given by (\ref{3.8}) - and hence the fourth order $S$ estimate -
once we specify $T[32]=\sum_{k=3}^5 \nu_{23.k}$ and $T[2^3]=\sum_{k=3}^6 \nu_{222.k}$.
Since $g_1 = \mu_2^{-1/2}$, $g_2 / \mu =g_{12} = -\mu_2^{-3/2}/2$, $g_{22} /\mu = g_{122} = 2 \mu_2^{-5/2}/4$,
and so on, these components are given by
\begin{eqnarray*}
 \nu_{23.3} &=&45 \beta_3 - 15 \beta (2\beta_3^2 +3 \beta_4 -3)/2,
\\
 \nu_{23.4} &=& 45 (3 \beta_5 -6 \beta_3 + 7 \beta_4-7)/4
\\
&&- 105 \beta \{ \beta_6 -3 \beta_4 +2 +12 \beta_3 (\beta_5 - 2 \beta_3)+6(\beta_4 -1)^2\}/16,
\\
\nu_{23.5} &=& 105 \{ 2 \beta_3 ( \beta_6-3  \beta_4+2)+3(\beta_4-1)
(\beta_5 -2 \beta_3) \} /16
\\
&&- 945 \beta ( \beta_4 -1) (\beta_6 -3  \beta_4 + 2 )/32,
\\
 \nu_{222.3} &=& -775 \beta,
\\
\nu_{222.4} &=& 45 \{ -y5 \beta_3 +7 \beta (\beta_4 -1+4 \beta_3^2) /4 \},
\\
\nu_{222.5} &=& 315 \{ -3 (\beta_4 -1) \beta_3 + \beta_3^3 + 3 \beta (\beta_4-1) (\beta_4 -1 + \beta_3^2)\} /16,
\\
\nu_{222.6} &=& 945 \{ -12 (\beta_4-1)^2 \beta_3 +11 \beta (\beta_4 -1)^3 \}/64.
\end{eqnarray*}

\noindent
[Figures 6.7 and 6.8 about here.]

It follows that a second order estimate of $\mu/\sqrt{\mu_2}$
is $\widehat{\mu}/\sqrt{\widehat{\mu_2}} - (1/n) \{ \widehat{\beta_3}/2 + (1/8) (\widehat{\mu}/\sqrt{\widehat{\mu_2}}) (3 \widehat{\beta_4} + 1) \}$.
Suppose now $Y = h (X)$ is normally distributed with mean $\mu$ and variance $\mu_2$.
Then a target estimator (Cabrera and Fernholz, 1999, 2004) of $\mu/\sqrt{\mu_2}$ is
$\sqrt{2} \Gamma ((n - 1)/2) \bar{Y}/\{ \sqrt{n - 1} \Gamma (n/2 - 1) S_Y \}$,
where $\bar{Y}$ and $S_Y$ denote the sample mean and the sample standard deviation, respectively.
Sen (1988)'s conditions for asymptotic normality do not hold here.
Figures 6.7 and 6.8 compare the performance of the two estimates and those obtained by bootstrapping and jackknifing.
Our estimate again provides the lowest absolute bias and the lowest mean squared error,
but the differences with respect to the target estimator do not appear to be significant.
\ $\Box$

\noindent
{\bf Example 6.6}
$k=s=1$, $T(F) =g (\mu_2, \mu_3) $ for $\mu_r = \mu_r (X)$, $X \sim F$ and $g$ having finite derivatives
\begin{eqnarray*}
g_{ij \cdots} = (\partial / \partial \mu_i) (\partial / \partial \mu_j) \cdots g(\mu_2, \mu_3).
\end{eqnarray*}
By Example 6.3,
$\mu_{3x} = \mu_x^3 - \mu_3 -3 \mu_2 \mu_x$,
$\mu_{3xy} = -3 (\mu_x + \mu_y) (\mu_x \mu_y - \mu_2)$,
$\mu_{3xyz} = 12 \mu_x \mu_y \mu_z$.
Now use (A8--A11) of Withers (1994a) with $U_i (F) = \mu_i$:
for $ T[2]$ we need $U_3 (a^2) = -6 \mu_3$, $U_{23} (a, a) = \mu_5 -4 \mu_3 \mu_2$,
$U_{33} (a, a) = \mu_6 - 6 \mu_4 \mu_2 - \mu_3^2 + 9 \mu_2^3$.
This gives
\begin{eqnarray*}
T[2] = -2 g_2 \mu_2 - 6 g_3 \mu_3 + g_{22} (\mu_4 - \mu_2^2) +
2 g_{23} (\mu_5-4 \mu_3 \mu_2) + g_{33} (\mu_6 -6 \mu_4 \mu_2 - \mu_3^2)
+ 9 \mu_2^3. \nonumber
\end{eqnarray*}
For $T[3]$ we need
\begin{eqnarray*}
U_3 (a^3) &=& 12 \mu_3,
\\
U_{23}  (a, a^2)  & =& -6 (\mu_5 - 2 \mu_3 \mu_2),
\\
U_{32} (a, a^2)& =& -2 (\mu_5 - 4 \mu_3 \mu_2),
\\
U_{33}  (a, a^2) &=& -6 (\mu_7 - 4 \mu_5 \mu_2)- \mu_4 \mu_3+4 \mu_3\mu_2^2,
\\
U_{223}  (a, a, a) &=& \mu_7 - 5 \mu_5 \mu_2 -\mu_4  \mu_3 + 8 \mu_3\mu_2^2,
\\
U_{233}  (a, a, a) &=& \mu_8 - 3 \mu_6 \mu_2 -2\mu_5  \mu_3 + 15 \mu_4\mu_2^2+ 4 \mu_3^2 \mu_2,
\\
 U_{333}  (a, a, a) &=& \mu_9 - 9 \mu_7 \mu_2 -3\mu_6 \mu_3 + 27 \mu_5\mu_2^2+ 18 \mu_4 \mu_3 \mu_2 + 2 \mu_3^3 -135 \mu_3 \mu_2^3.
\end{eqnarray*}
This gives
\begin{eqnarray*}
 T[3] &=& 12 g_3 \mu_3 - 6 g_{22} (\mu_4 - \mu_2^2) -12 g_{23}
(2 \mu_5 - 5 \mu_3 \mu_2)
\\
&& - 18 g_{33} (\mu_7 - 4 \mu_5 \mu_2 - \mu_4 \mu_3 + 4 \mu_3 \mu_2^2)
\\
&&+ g_{222} (\mu_6 -3 \mu_4 \mu_2 + 2 \mu_2^3) + 3 g_{223} (\mu_7 -5 \mu_5
\mu_2 - \mu_4 \mu_3 + 8 \mu_3 \mu_2^2 )
\\
&&+ g_{233} U_{233} (a, a, a) + g_{333} U_{333} (a, a, a).
\end{eqnarray*}
For $T[2^2] $ we need
\begin{eqnarray*}
U_{23} (a, ab^2) &=& 12 \mu_3 \mu_2,
\\
U_{33} (a, ab^2)& =& 12 (\mu_4-3 \mu_2^2) \mu_2,
\\
U_{23} (ab, ab)  &=& 12 \mu_3 \mu_2,
\\
U_{33} (ab, ab) &=& 18 (\mu_4 \mu_2 +\mu_3^2 - \mu_2^3),
\end{eqnarray*}
and at $(ab, a, b)$:
\begin{eqnarray*}
U_{222} &=& -2 \mu_3^2,
\\
U_{223} &=& U_{232} = -2 \mu_3 \kappa_4,
\\
U_{233} &=& -2 \kappa_4^2,
\\
 U_{322}& =&-6 \mu_3 (\mu_4-\mu_2^2),
\\
U_{323} &=& U_{332} =-3 \{ (\mu_4-\mu_2^2) \kappa_4+\mu_3(\mu_5-4 \mu_3 \mu_2)\},
\\
U_{333} &=& -6 \kappa_4 (\mu_5 - 4 \mu_3 \mu_2).
\end{eqnarray*}
This gives
\begin{eqnarray*}
 T[2^2] &=& 12 g_{22} \sigma^4 + 48 g_{23} \sigma^5 \beta_3 + 12
g_{33} \sigma^6 (7 \beta_4 + 6 \beta_3^2 - 15)
\\
&& - 8 g_{222} \sigma^6 (\beta_4 -1+\beta_3^2)-4 g_{223} \sigma^7 (2 \beta_5
+13 \beta_4 \beta_3 -29 \beta_3)
\\
&& - 4 g_{233} \sigma^8 ( \beta_6 + 12 \beta_5 \beta_3 +2 \beta_4^2 - 42
\beta_4 -49 \beta_3^2+9)
\\
&& - 12 g_{333} \sigma^9 \{(\beta_6 - 14 \beta_4 - \beta_3^2 +33)\beta_3
+2 (\beta_4-3)\beta_5\}
\\
&& + g_{2222} \sigma^8 (\beta_4 -1)^2 + 4 g_{2223} \sigma^9
(\beta_5 - 4 \beta_3) (\beta_4 -1)
\\
&&+ 2 g_{2233} \sigma^{10} \{ ( \beta_6 - 6 \beta_4 - \beta_3^2) (\beta_4-1)
+2 (\beta_5 - 4 \beta_3)^2\}
\\
&&+4 g_{2333} \sigma^{11} (\beta_5 -4 \beta_3)(\beta_6 -6 \beta_4 - \beta_3^2)
+g_{3333} \sigma^{12}(\beta_6 - 6 \beta_4 - \beta_3^2)^2,
\end{eqnarray*}
where again $\beta_r = \sigma^{-r} \mu_r$.
By (\ref{3.6}), (\ref{3.7}) this gives $S_1(F)$ and $S_2 (F)$ so that (\ref{6.1}) gives the third order $S$ estimate. \ $\Box$

\noindent
{\bf Example 6.6.1}
$g(\mu_2, \mu_3) = \mu_3 \mu_2^{-3/2}$, the skewness of $F$.
One obtains
\begin{eqnarray*}
& T[2] &= -3 \beta_5 +15 (\beta_4-1) \beta_3 /4 +9 \beta_3,
\\
& T[3] &= -105 (\beta_6 -3 \beta_4 +2) \beta_3/8 +36 \beta_5 - 45 \beta_4
\beta_3 /2 - 111 \beta_3/2,
\\
& T[2^2] &= -105 (\beta_5 -4 \beta_3) (\beta_4 -1)/2 +945
(\beta_4 -1)^2 \beta_3 /16 + \beta_3 \delta,
\end{eqnarray*}
where $\delta = 78-105 (\beta_4+\beta_3^2)-15 (2 \beta_5+13 \beta_4 \beta_3 -29  \beta_3)$.
Note that T[2] for both skewness and  kurtosis were first given in Withers (1994b):
$\beta_i$ in (\ref{3.2}) there should be $\beta_{i+r}$.
$\Box$

\section*{Acknowledgments}

The authors would like to thank the Editor--in--Chief (Professor Pranab K. Sen),
the Co--Editor and the referee
for carefully reading the paper and for their comments which greatly improved the paper.

\newpage

\section*{Appendix A}

Here we explain the infinitesimal jackknife given in the unpublished report
Jaeckel (1972) and its relation to our case $p=2, k=1$.
In fact he offers two forms and these are equivalent to
our second order $S$ and $T$ estimates, but are arrived at without using
functional derivatives.

He considers the case $k=1$ for $T(F)$ but with $\widehat{F(x)}$ the {\it weighted}
empirical distribution, $\widehat{F(x)} = \sum_{i=1}^n w_i I(X_i \leq x)$.
His first form is  equivalent to our $T$ estimate with
\begin{eqnarray*}
T_1(\widehat{F}) = -(1/2) n^{-1} \sum_{i=1}^n\partial^2 T(\widehat{F}) /\partial w_i^2
\end{eqnarray*}
evaluated at $w_1= \cdots =w_n=1/n$.
His second form is essentially our $S$ estimate and is defined in the same way
except that in the definition of the weighted empirical distribution $ w_i $ is replaced by $w_i/(\sum_{j=1}^n w_j)$.

\newpage

\section*{Appendix B}

The following procedures in MAPLE calculate the $V$, $S$ and $T$
estimates described in Sections 2 to 4 to obtain bias of any given order.
The electronic versions of these procedures can be obtained from the corresponding author.

\begin{verbatim}
#this procedure expresses the first u symmetric polynomials in terms of power
#sum functions the output from this procedure is needed as input for
#the Vest() and Vestsum() procedures
symmpoly := proc(u)
local m,e;
e[0]:=1;
for m from 1 to u do
e[m]:=(1/m)*sum((-1)**(i-1)*e[m-i]*p[i],i=1..m);
od;
return([seq(e[mm],mm=1..u)]);
end proc;


#this procedure computes the G function given by (4.1)
gbetai := proc (beta,i)
local tt,ttt,ii,j,k,p;
tt:=0;
for ii from 1 to i do
p:=composition(beta,ii);
for j from 1 to nops(p) do
ttt:=1;
for k from 1 to ii do
ttt:=ttt*k**(p[j][k]);
od;
tt:=tt+ttt;
od;
od;
return(tt);
end proc;


#this procedure computes the D function given by (4.2)
dalphai := proc (alpha,i)
gbetai(alpha-i,i);
end proc;


#this procedure computes the coefficients c[k,j,r] given by equation (3.1)
cc := proc (k, j, r)
local d,tt,i,n;
d[j]:=1/Amkr(j,j,r);
tt:=d[j];
if (j>k) then
for i from 1 to (j-k) do
d[j-i]:=0;
for n from (j-i+1) to j do
d[j-i]:=d[j-i]+d[n]*Amkr(j-i,n,r);
od;
d[j-i]:=-d[j-i]/(Amkr(j-i,j-i,r));
od;
tt:=d[k];
end if;
return(tt);
end proc;


#the following procedure is needed as part of the above cc (k, j, r) procedure
Amkr := proc (m, k, r)
sum(((-1)**(l+k))*stirling1(k,l)*binomial(l,m)*(1-r)**(l-m),l=m..k);
end proc;


#this procedure computes the k (ndim) fold summation in equation (2.9)
#here it is assumed implicitly that k=5
#but the procedure can be applied for any k by making appropriate changes to
#the five do-loops and the five statements following it
#this procedure requires as input: n[1],n[2],...,n[k],r[1],r[2],...,r[k],
#ndim=the value of k,the output from symmpoly() and the T(.) function
#in equation (2.4) the output from this procedure is needed as input for
#the Vest() and coefficients() procedures
Vestsum := proc (ndim,n::list,r::list,e::list,T)
local kk,pa,ss,i1,i2,i3,i4,i5,paa,bb,tt,kkk,j,nn,ttt,mmm;
for kk from 1 to ndim do
pa[kk]:=partition(r[kk]);
od;
ss:=0;
for i1 from 1 to nops(pa[1]) do
for i2 from 1 to nops(pa[2]) do
for i3 from 1 to nops(pa[3]) do
for i4 from 1 to nops(pa[4]) do
for i5 from 1 to nops(pa[5]) do
paa[1]:=pa[1][i1];
paa[2]:=pa[2][i2];
paa[3]:=pa[3][i3];
paa[4]:=pa[4][i4];
paa[5]:=pa[5][i5];
bb:=member(1,paa[1])=false;
for kk from 2 to ndim do
bb:=bb and member(1,paa[kk])=false;
od;
if (bb) then
for kk from 1 to ndim do
tt[kk]:=expand(e[r[kk]]);
od;
for kkk from 1 to ndim do
for j from 1 to r[kkk] do
nn:=0;
for kk from 1 to nops(paa[kkk]) do
if (j=paa[kkk][kk]) then nn:=nn+1 end if;
od;
if (nn>0) then tt[kkk]:=coeff(tt[kkk],p[j],nn) end if;
od;
od;
ttt:=1;
for mmm from 1 to ndim do
ttt:=ttt*tt[mmm]*(n[mmm])**(nops(paa[mmm]));
od;
ss:=ss+ttt*T([seq(paa[mm],mm=1..ndim)]);
end if;
od;
od;
od;
od;
od;
ss:=expand(ss);
return(ss);
end proc;


#this procedure computes of the coefficients of the output from
#with respect to powers of n[1],n[2],...,n[k]
#It is assumed implicitly that k is 5
#the procedure can be applied to any k by changing the number
#of do loops and making other appropriate changes
#this procedure requires as input: n[1],n[2],...,n[k],r[1],r[2],...,r[k]
#and the output from Vestsum()
#the output from this procedure is needed as input for the Sest()
#and Test() procedures
coefficients := proc(n::list,r::list,ss)
local i1,i2,i3,i4,i5,sss,A;
A:=array(1..r[1]/2,1..r[2]/2,1..r[3]/2,1..r[4]/2,1..r[5]/2);
for i1 from 1 to r[1]/2 do
for i2 from 1 to r[2]/2 do
for i3 from 1 to r[3]/2 do
for i4 from 1 to r[4]/2 do
for i5 from 1 to r[5]/2 do
sss:=coeff(ss,n[1],i1);
sss:=coeff(sss,n[2],i2);
sss:=coeff(sss,n[3],i3);
sss:=coeff(sss,n[4],i4);
sss:=coeff(sss,n[5],i5);
A[i1,i2,i3,i4,i5]:=sss;
od;
od;
od;
od;
od;
return(A);
end proc;


#this procedure computes V_r for the V estimate in Section 2
#this procedure requires as input: n[1],n[2],...,n[k],r[1],r[2],...,r[k],
#ndim=the value of k,the output from symmpoly() and the T(.) function
#in equation (2.4)
Vest := proc(ndim,n::list,r::list,e::list,T)
local ss,mm;
ss:=Vestsum(ndim,n,r,e,T);
ss:=ss*(product(factorial(n[mm]-r[mm]),mm=1..ndim));
ss:=ss/(product(factorial(n[mm]),mm=1..ndim));
ss:=simplify(ss);
return(ss);
end proc;


#this procedure computes V_r for the S estimate in Section 3
#It is assumed implicitly that k is 5
#the procedure can be applied to any k by changing the number
#of do loops and making other appropriate changes
#this procedure requires as input: n[1],n[2],...,n[k],
#r[1],r[2],...,r[k] and the output from coefficients()
Sest := proc(n::list,r::list,A::array)
local i1,i2,i3,i4,i5,j1,j2,j3,j4,j5,tt,ttt,ss;
ss:=0;
for i1 from r[1]/2 to r[1]-1 do
for i2 from r[2]/2 to r[2]-1 do
for i3 from r[3]/2 to r[3]-1 do
for i4 from r[4]/2 to r[4]-1 do
for i5 from r[5]/2 to r[5]-1 do
tt:=0;
for j1 from r[1]-i1-1 to r[1]/2-1 do
for j2 from r[2]-i2-1 to r[2]/2-1 do
for j3 from r[3]-i3-1 to r[3]/2-1 do
for j4 from r[4]-i4-1 to r[4]/2-1 do
for j5 from r[5]-i5-1 to r[5]/2-1 do
ttt:=A[j1+1,j2+1,j3+1,j4+1,j5+1]*cc(r[1]-i1-1,j1,r[1])*cc(r[2]-i2-1,j2,r[2]);
ttt:=ttt*cc(r[3]-i3-1,j3,r[3])*cc(r[4]-i4-1,j4,r[4])*cc(r[5]-i5-1,j5,r[5]);
tt:=tt+ttt;
od;
od;
od;
od;
od;
tt:=tt*factorial(n[1]-i1-1)*factorial(n[2]-i2-1)*factorial(n[3]-i3-1);
tt:=tt*factorial(n[4]-i4-1)*factorial(n[5]-i5-1);
tt:=tt/(factorial(n[1]-1)*factorial(n[2]-1));
tt:=tt/(factorial(n[3]-1)*factorial(n[4]-1)*factorial(n[5]-1));
ss:=ss+tt;
print(i1,i2,i3,i4,i5,ss);
od;
od;
od;
od;
od;
ss:=simplify(ss);
return(ss);
end proc;


#this procedure computes V_r for the T estimate in Section 4
#It is assumed implicitly that k is 5
#the procedure can be applied to any k (ndim) by changing the number
#of do loops and making other appropriate changes
#this procedure requires as input: n[1],n[2],...,n[k],r[1],r[2],...,r[k],
#ndim=the value of k and the output from coefficients()
Test := proc(ndim,n::list,r::list,A::array)
local i1,i2,i3,i4,i5,j1,j2,j3,j4,j5,nn,tt,ttt,tttt,ss,a,kk;
ss:=0;
for i1 from r[1]/2 to r[1]-1 do
for i2 from r[2]/2 to r[2]-1 do
for i3 from r[3]/2 to r[3]-1 do
for i4 from r[4]/2 to r[4]-1 do
for i5 from r[5]/2 to r[5]-1 do
tt:=0;
for j1 from r[1]-i1-1 to r[1]/2-1 do
for j2 from r[2]-i2-1 to r[2]/2-1 do
for j3 from r[3]-i3-1 to r[3]/2-1 do
for j4 from r[4]-i4-1 to r[4]/2-1 do
for j5 from r[5]-i5-1 to r[5]/2-1 do
ttt:=A[j1+1,j2+1,j3+1,j4+1,j5+1]*cc(r[1]-i1-1,j1,r[1])*cc(r[2]-i2-1,j2,r[2]);
ttt:=ttt*cc(r[3]-i3-1,j3,r[3])*cc(r[4]-i4-1,j4,r[4])*cc(r[5]-i5-1,j5,r[5]);
tt:=tt+ttt;
od;
od;
od;
od;
od;
nn[1]:=i1;
nn[2]:=i2;
nn[3]:=i3;
nn[4]:=i4;
nn[5]:=i5;
ttt:=1;
for kk from 1 to ndim do
tttt:=0;
for a from nn[kk] to 100 do
tttt:=tttt+(n[kk])**(-a)*dalphai(a,nn[kk]);
od;
ttt:=ttt*tttt;
od;
ss:=ss+tt*ttt;
print(i1,i2,i3,i4,i5,ss);
od;
od;
od;
od;
od;
ss:=simplify(ss);
return(ss);
end proc;
\end{verbatim}

\newpage

\begin{center}
{\bf Table 5.1.}~~Number of terms for estimates of bias $O(n^{-p})$ for $k=1$.\\
\begin{tabular}{|l|l|l|l|l|l|l|l|} \hline
$p=$ & 1& 2& 3& 4& 5& 6&7\\ \hline
$V_{\pmb {n} p}(\widehat{F})$ & 1& 2& 5&11&22 &42&77\\
$S_{\pmb {n} p}(\widehat{F})$ & 1& 2& 4&8&15 &25&44\\
$T_{\pmb {n} p}(\widehat{F})$ & 1& 2& 5&11&21 &$\cdot$&$\cdot$\\ \hline
\end{tabular}
\end{center}

\begin{center}
{\bf Table 5.2.}~~Number of terms for estimates of bias $O(n_{0}^{-p})$, where $n_{0}$ = min $n_{i}$.\\
{\small\renewcommand{\tabcolsep}{1pt}
\begin{tabular}{|c|c|c|c|c|c|} \hline
$p=$ & 1& 2& 3& 4 & 5 \\ \hline
$V_{\pmb {n} p}(\widehat{F})$ & $1$ & $k+1$ & $s_2+4k+1$ & $s_3 +8s_2 +10k+1$ & $s_{4}+14s_{3}+31s_{2}+21k+1$ \\
& & & $=(k^2 +7k+2)/2$ & $=(k^3+21k^2 +38k+6)/6$ & $=(k^{4}+40k^{3}+305k^{2}+230k+24)/24$\\
$S_{\pmb {n} p}(\widehat{F})$ & 1& $k+1$ & $s_{2} + 3k+1$ & $s_{3}+3s_{2}+7k+1$ & $s_{4}+9s_{3}+15s_{2}+14k+1$\\
& & & $=(k^{2}+5k+2)/2$ & $=(k^{3}+6k^{2}+35k+6)/6$ & $=(k^{4}+30k^{3}+143k^{2}+234k+24)/24$\\
$ T_{\pmb {n} p}(\widehat{F})$ & $1$ & $k+1$ & $s_{2}+4k+1$ & $ s_{3}+7s_{2}+10k +1$ & $s_{4}+13s_{3}+25 s_{2}+20k+1$\\
& & & $=(k^{2}+7k+2)/2$ & $=(k^{3}+18k^{2}+20k+6)/6$ & $=(k^{4}+36k^{3}+245k+270k+24)/24$\\ \hline
\end{tabular}}
\end{center}

\begin{center}
{\bf Table 5.3.}~~Number of terms for estimates of bias $O(n^{-p})$ for $k=2$ and 3.\\
\begin{tabular}{|c|c|c|r|r|r||c|c|r|r|r|}
\hline
\multicolumn{6}{|c||}{$k=2$} & \multicolumn{5}{c|}{$k = 3$} \\ \hline
$p=$ & 1 & 2 & 3  & 4  & 5  & 1 & 2 & 3  & 4  & 5 \\ \hline
$V_{\pmb {n}p} (\widehat{F})$  & 1 & 3 & 10 & 29 & 74 & 1 & 4 & 16 & 56 & 171 \\
$S_{\pmb {n}p} (\widehat{F})$  & 1 & 3 & 8  & 18 & 44 & 1 & 4 & 13 & 32 & 97\\
$T_{\pmb {n}p} (\widehat{F})$  & 1 & 3 & 10 & 28 & 66 & 1 & 4 & 16 & 53 & 149\\ \hline
\end{tabular}
\end{center}

\newpage

\centerline{\epsfig{file=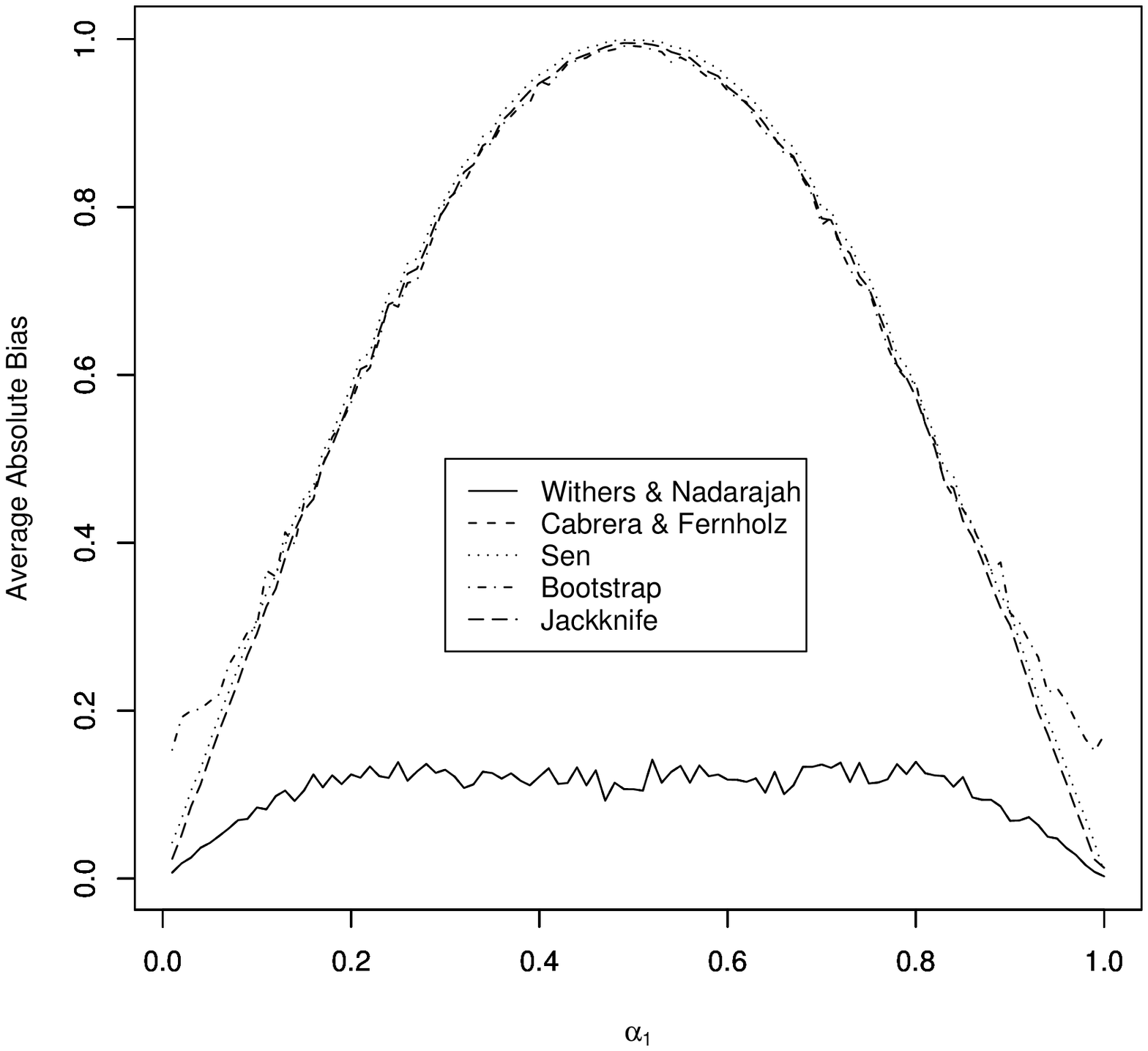,width=6in,height=3.8in}}
\noindent
{\bf Figure 6.1.}~~The average absolute bias of the estimator of $\{ \alpha_1 \mu_1 + (1 - \alpha_1) \mu_2 \}^2$
when $\mu_1$ and $\mu_2$ are the means of two independent normal distributions with unit standard deviations.
The average is based on 100 random samples each of size 100.

\centerline{\epsfig{file=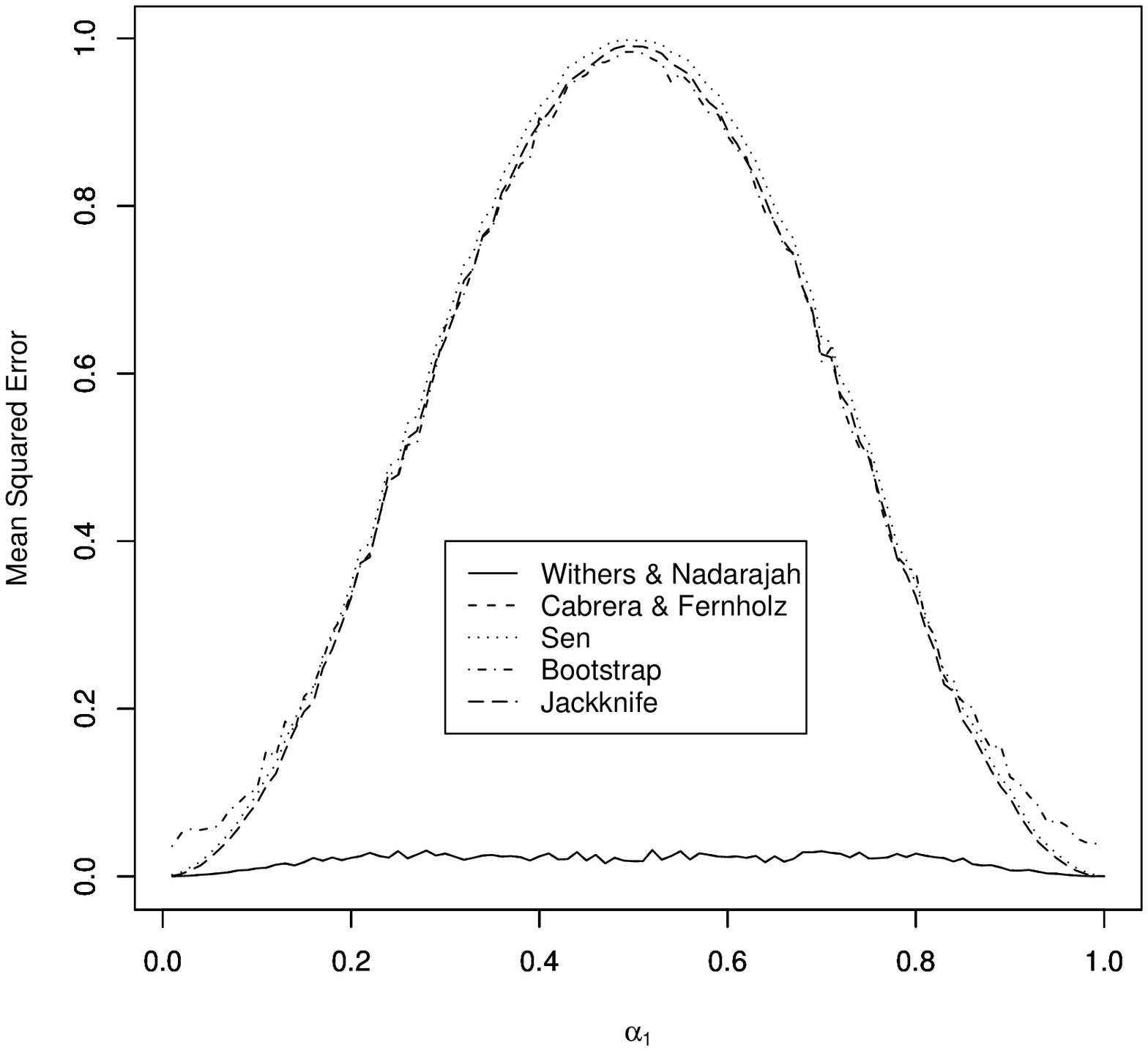,width=6in,height=3.8in}}
\noindent
{\bf Figure 6.2.}~~Mean squared error of the estimator of $\{ \alpha_1 \mu_1 + (1 - \alpha_1) \mu_2 \}^2$
when $\mu_1$ and $\mu_2$ are the means of two independent normal distributions with unit standard deviations.
The MSE is based on 100 random samples each of size 100.

\centerline{\epsfig{file=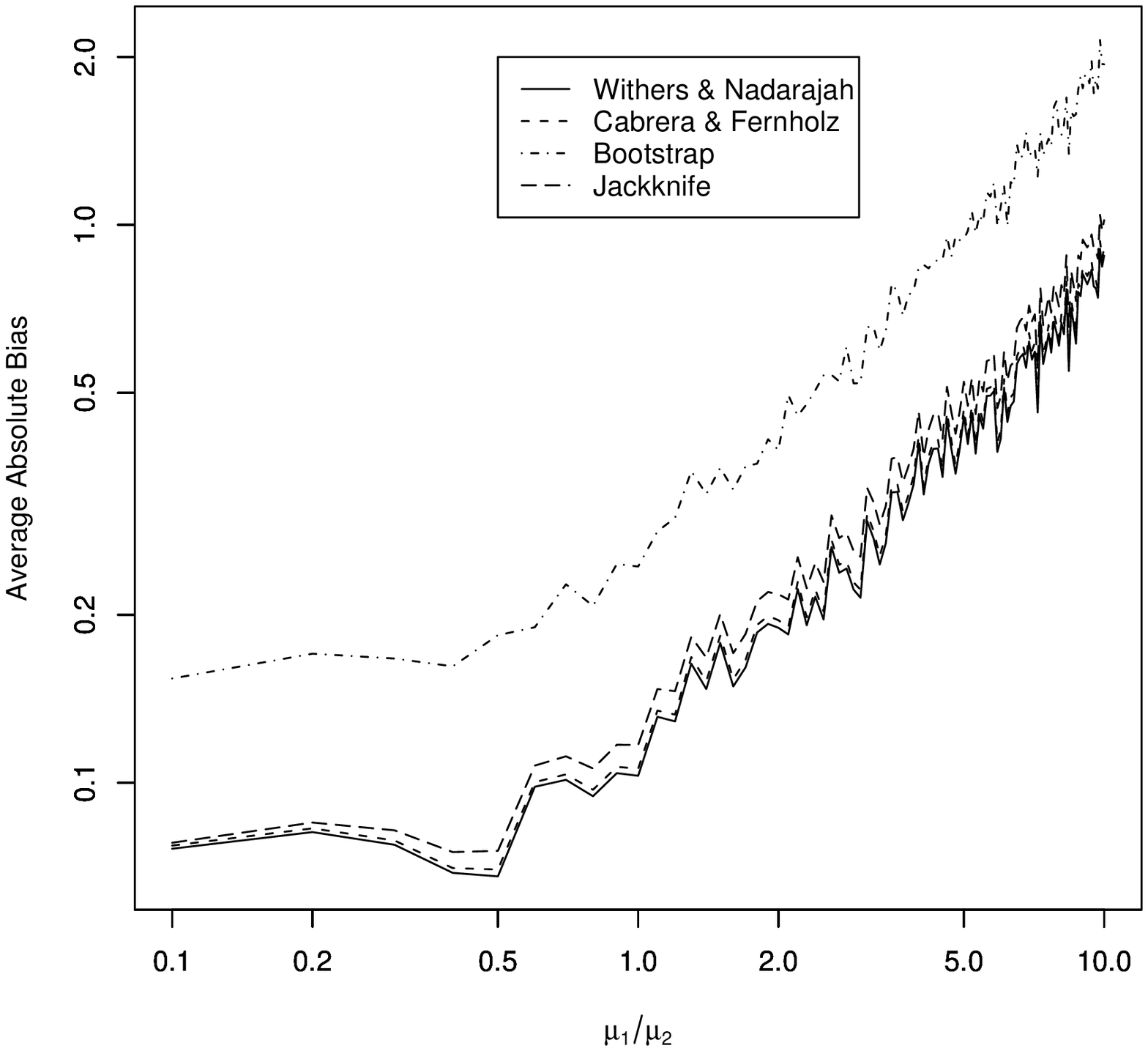,width=6in,height=3.8in}}
\noindent
{\bf Figure 6.3.}~~The average absolute bias of the estimator of $\mu_1/\mu_2$
when $\mu_1$ and $\mu_2$ are the means of two independent exponential distributions.
The average is based on 100 random samples each of size 100.
The $x$ axis is in log scale.

\centerline{\epsfig{file=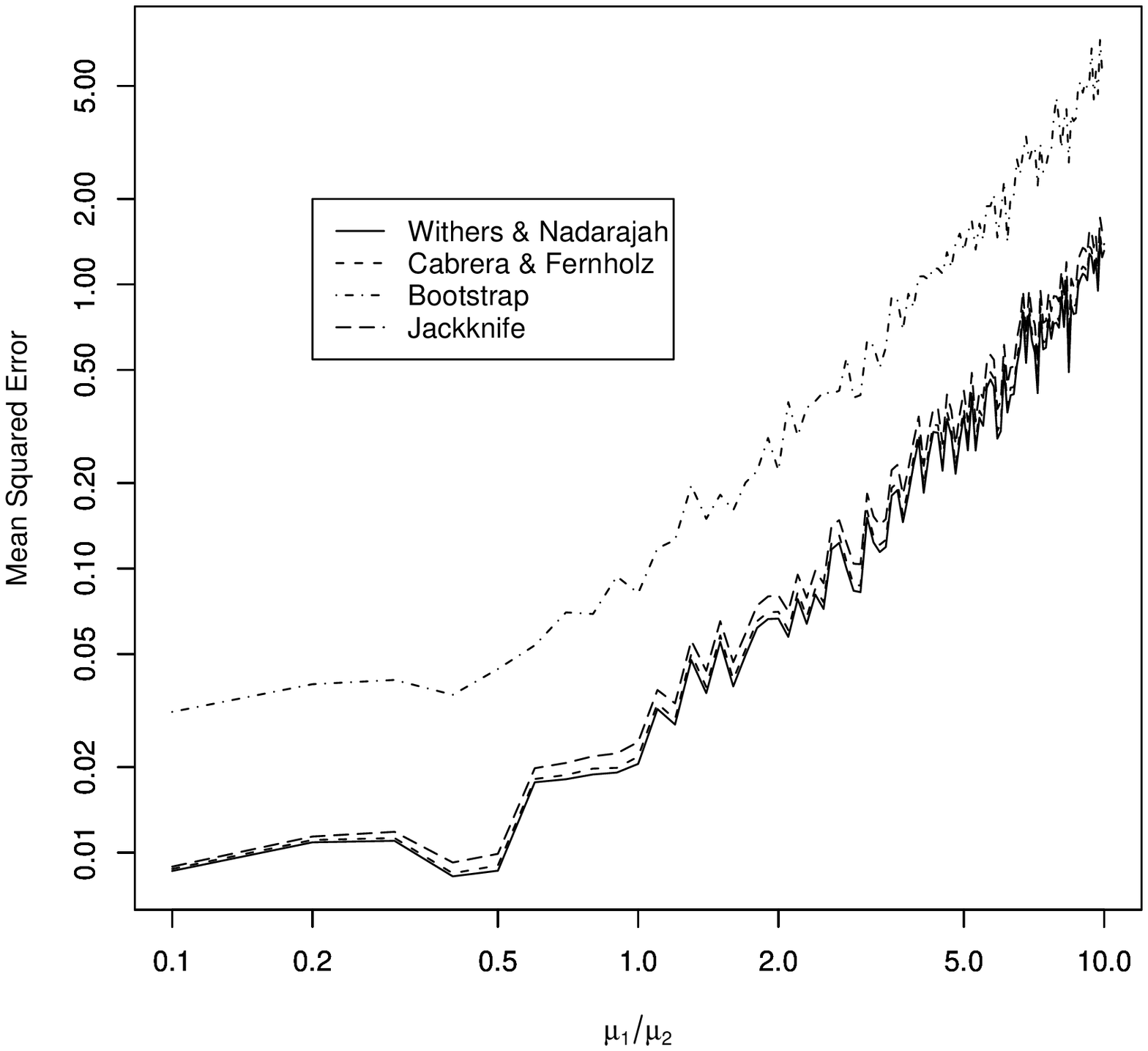,width=6in,height=3.8in}}
\noindent
{\bf Figure 6.4.}~~Mean squared error of the estimator of $\mu_1/\mu_2$
when $\mu_1$ and $\mu_2$ are the means of two independent exponential distributions.
The MSE is based on 100 random samples each of size 100.
The $x$ axis is in log scale.

\centerline{\epsfig{file=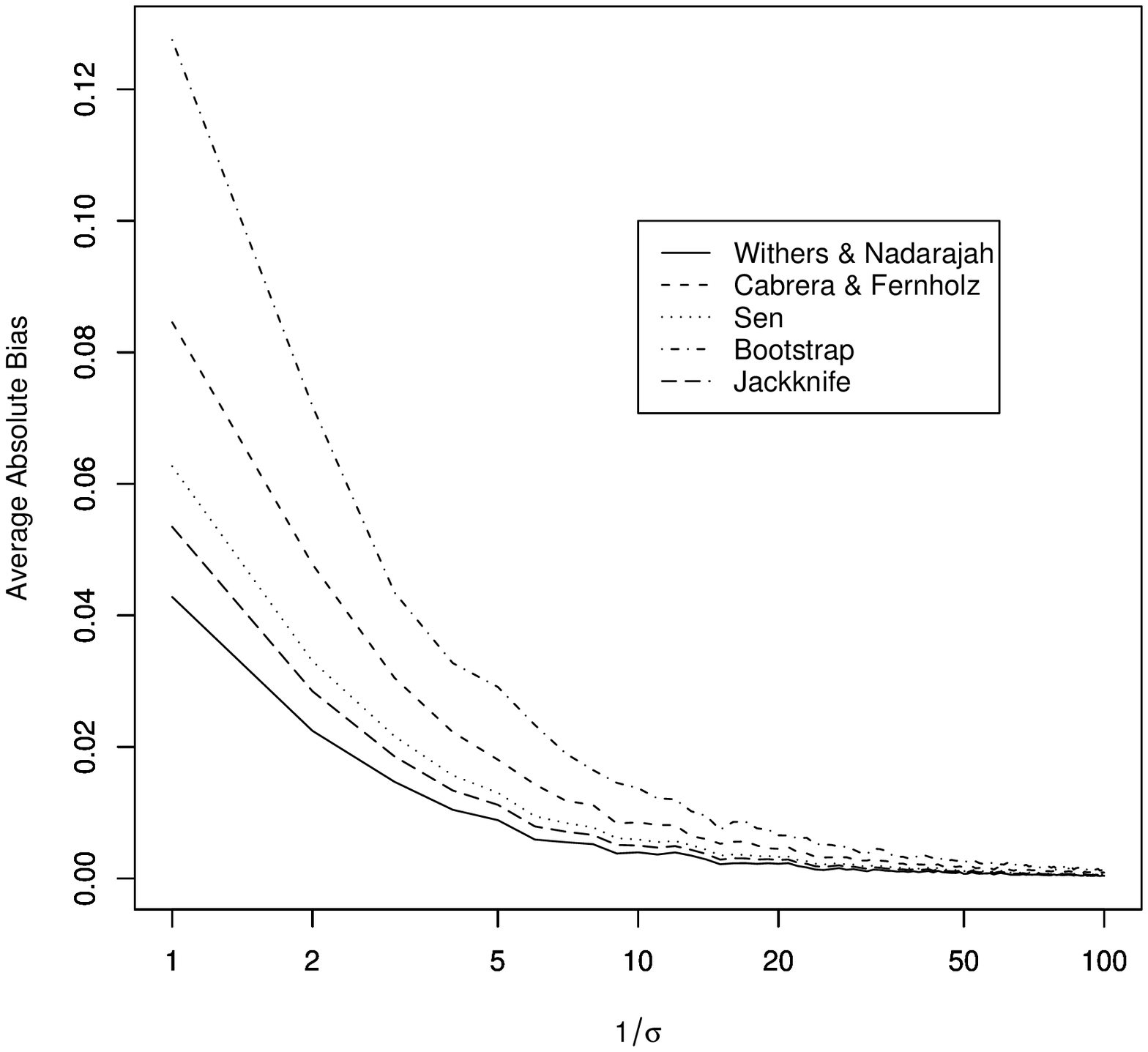,width=6in,height=3.8in}}
\noindent
{\bf Figure 6.5.}~~The average absolute bias of the estimator of $\sigma$
when $\sigma$ is the mean of an exponential distribution.
The average is based on 100 random samples each of size 100.
The $x$ axis is in log scale.

\centerline{\epsfig{file=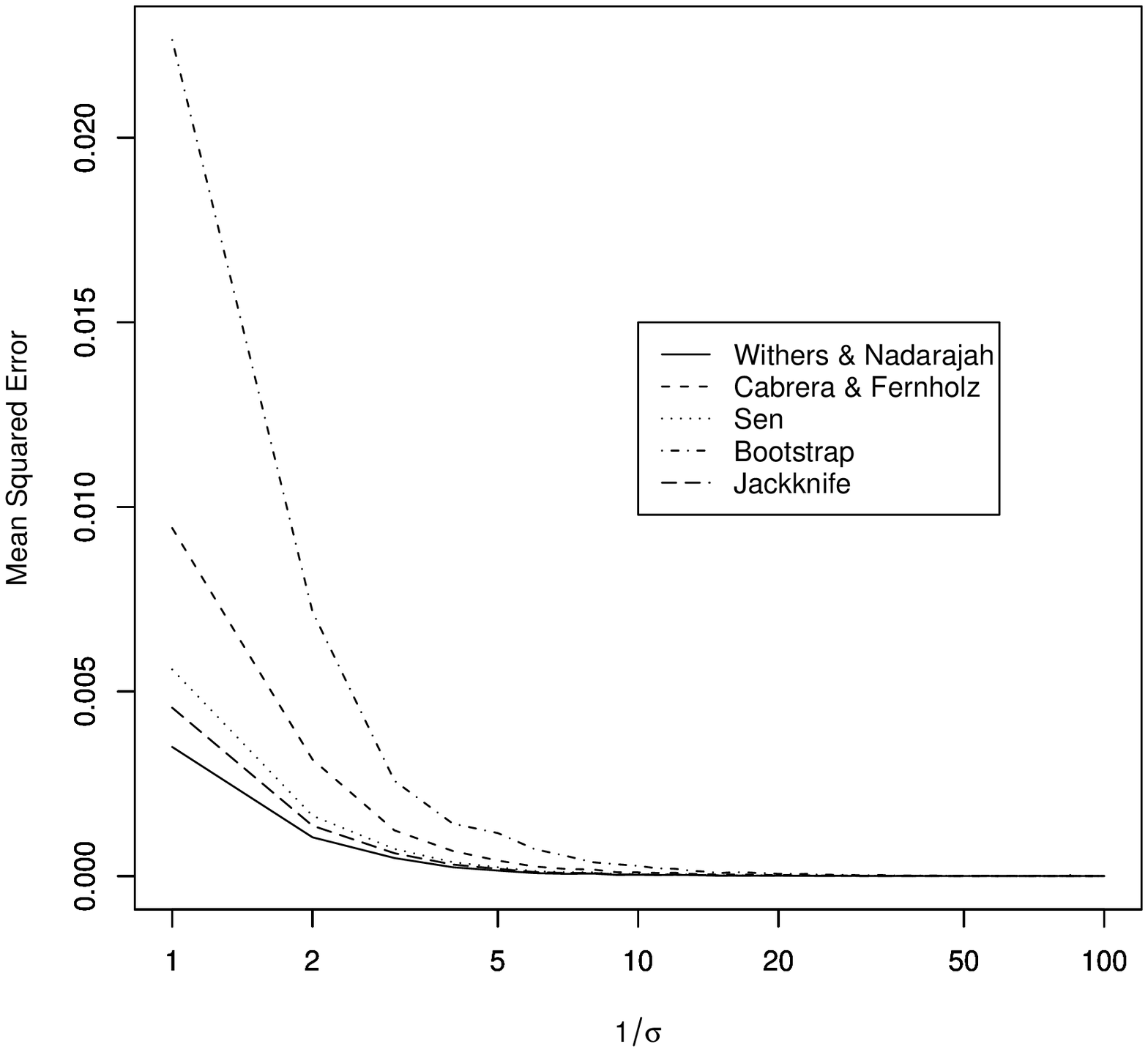,width=6in,height=3.8in}}
\noindent
{\bf Figure 6.6.}~~Mean squared error of the estimator of $\sigma$
when $\sigma$ is the mean of an exponential distribution.
The average is based on 100 random samples each of size 100.
The $x$ axis is in log scale.

\centerline{\epsfig{file=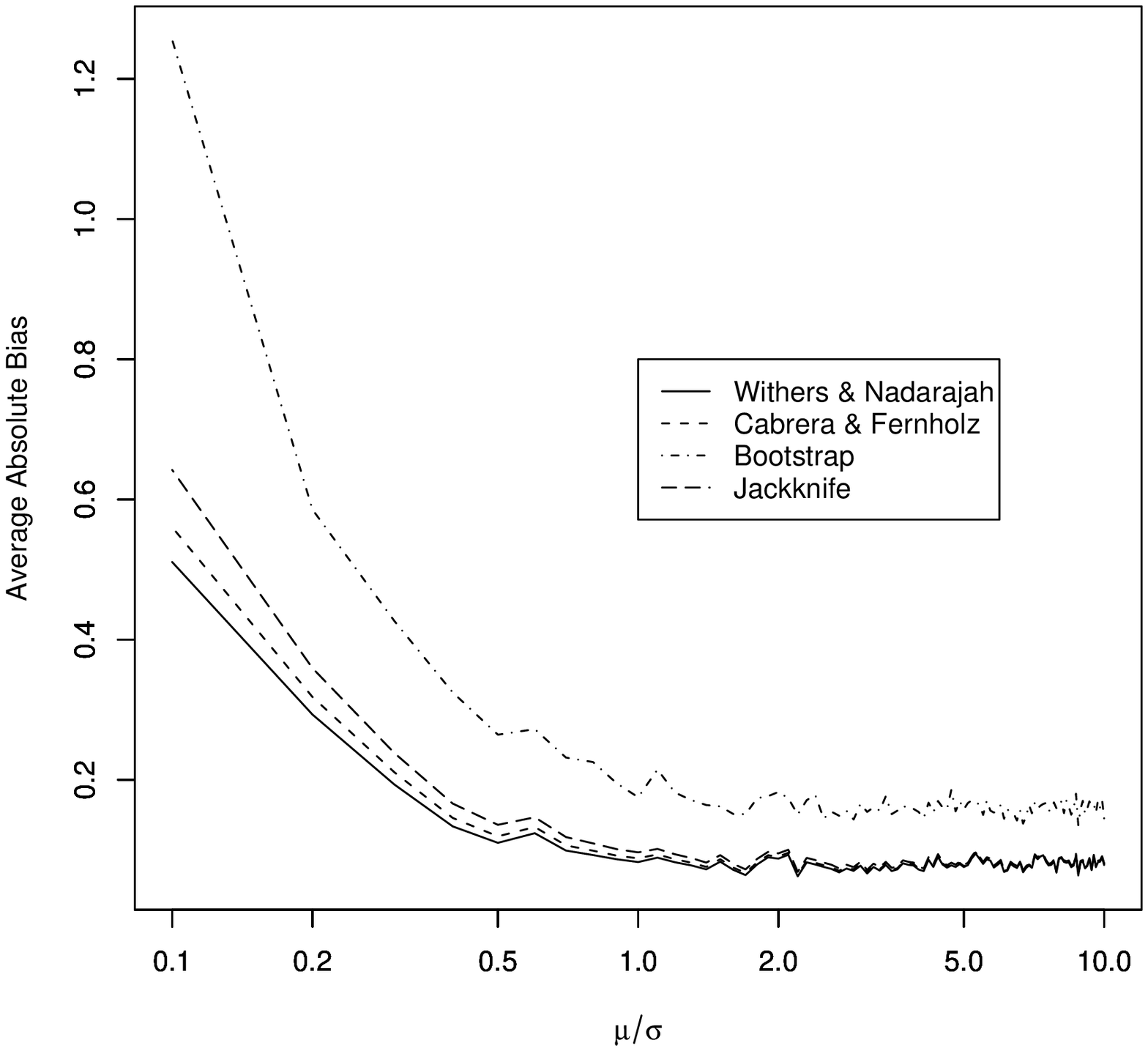,width=6in,height=3.8in}}
\noindent
{\bf Figure 6.7.}~~The average absolute bias of the estimator of $\mu/\sigma$
when $\mu$ and $\sigma$ are the mean and the standard deviation of a normal distribution.
The average is based on 100 random samples each of size 100.
The $x$ axis is in log scale.

\centerline{\epsfig{file=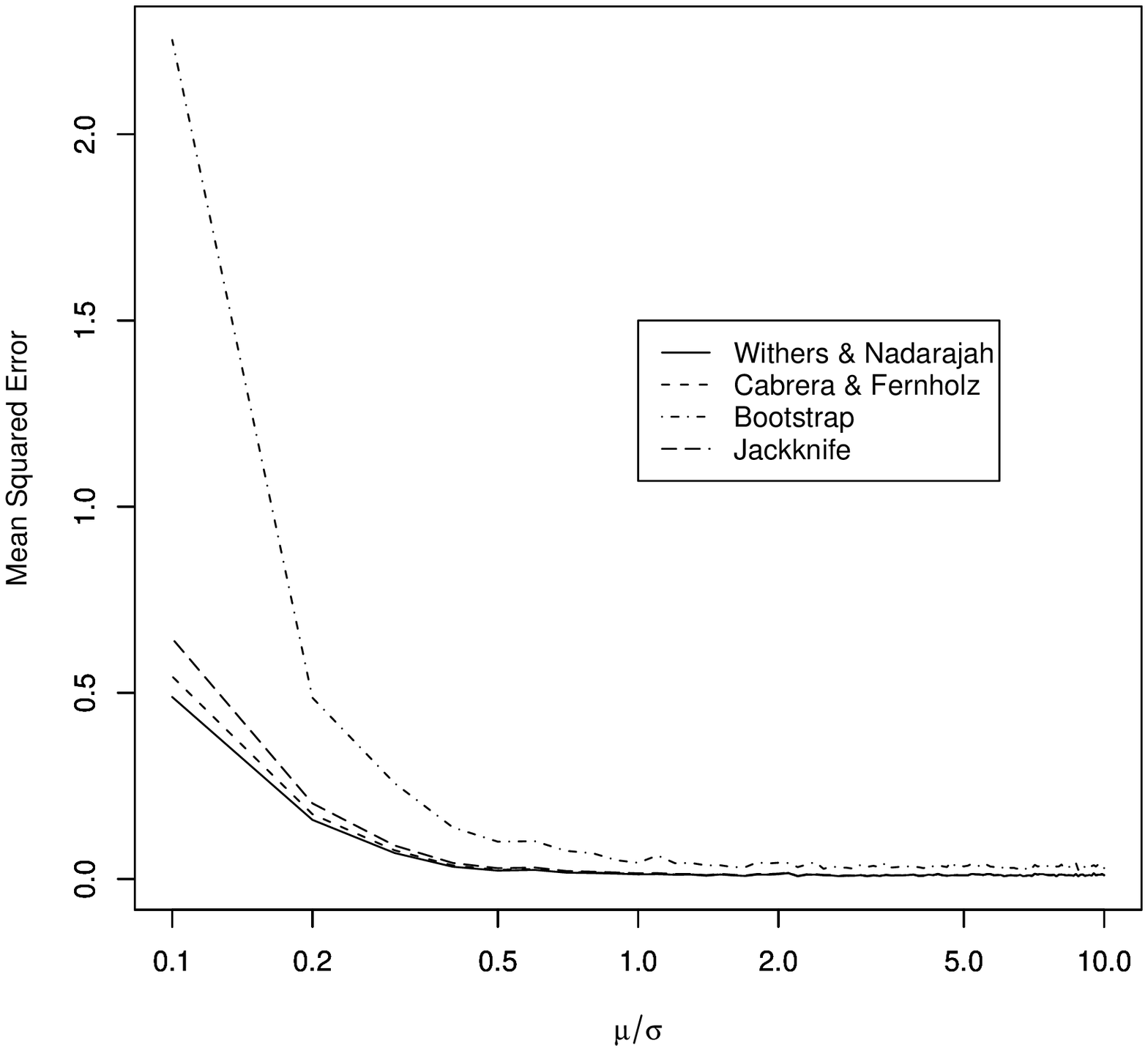,width=6in,height=3.8in}}
\noindent
{\bf Figure 6.8.}~~The mean squared error of the estimator of $\mu/\sigma$
when $\mu$ and $\sigma$ are the mean and the standard deviation of a normal distribution.
The MSE is based on 100 random samples each of size 100.
The $x$ axis is in log scale.

\end{document}